\documentclass[12pt]{article}
\usepackage{epsfig}
\usepackage{graphicx}
\usepackage{a4}
\usepackage{latexsym}
\usepackage{cite}

\usepackage{color}
\usepackage{colordvi}

\graphicspath{{Pics/}}
\DeclareGraphicsExtensions{.eps,.ps}

\usepackage{pslatex}
\usepackage[latin1]{inputenc}
\usepackage[T1]{fontenc}

\def\nc{\ca}
\def\ca{{C^{}_A}}
\def\cas{{C^{\,2}_A}}
\def\cf{{C^{}_F}}
\def\cfs{{C^{\,2}_F}}
\def\nf{{n^{}_{\! f}}}

\def\mt{{m_t}}
\def\shad{{s_{\rm had}}}
\def\shat{{\hat s}}
\def\muf{{\mu^{}_f}}
\def\mufs{{\mu^{\,2}_f}}
\def\mur{{\mu^{}_r}}
\def\murs{{\mu^{\,2}_r}}
\def\alphas{{\alpha_s}}

\def\z#1{{\zeta_{#1}}}
\def\lntwo{{\ln 2}}
\def\lntwos{{\ln^{\,2} 2}}
\def\lntwoc{{\ln^{\,3} 2}}
\def\lntwof{{\ln^{\,4} 2}}

\def\lnN{\ln \tilde{N}}
\def\lnNs{\ln^{\,2} \tilde{N}}
\def\lnNc{\ln^{\,3} \tilde{N}}
\def\lnNf{\ln^{\,4} \tilde{N}}

\def\lnbeta{\ln \beta}
\def\lnbetas{\ln^{\,2} \beta}
\def\lnbetac{\ln^{\,3} \beta}
\def\lnbetaf{\ln^{\,4} \beta}

\def\NNLO{{\mbox{NNLO (approx)}}}
\def\sigmaNNLO{\ensuremath{{\sigma_{\scriptstyle\NNLO}}}}

\def\NLO{{\mbox{NLO}}}
\def\sigmaNLO{\ensuremath{{\sigma_{\scriptstyle\NLO}}}}
\def\sigmaRES{\ensuremath{{\sigma_{\scriptstyle \mbox{res}}}}}

\def\MSbar{\ensuremath{\overline{\mbox{MS}}}}

\newcommand{\lsim}{\raisebox{-0.07cm}{$\:\stackrel{<}{{\scriptstyle \sim}}\: $} }

\newcommand{\ec}{\gamma_e}
\newcommand{\ecs}{\gamma_e^{\,2}}
\newcommand{\ect}{\gamma_e^{\,3}}
\newcommand{\ecf}{\gamma_e^{\,4}}
\def\m2{m^2}
\def\b#1{\beta_#1}
\def\Aq#1{A^{(#1)}_q}
\def\Dq#1{D^{(#1)}_q}
\def\DQQ#1{D^{(#1)}_{Q{\bar Q}}}
\newcommand{\Lqr}{\ln (4m^2/\mu_r^{\,2})}
\newcommand{\Lqrs}{\ln^2 (4m^2/\mu_r^{\,2})}
\newcommand{\Lfr}{\ln (\mu_f^{\,2}/\mu_r^{\,2})}
\newcommand{\Lfrs}{\ln^2 (\mu_f^{\,2}/\mu_r^{\,2})}

\begin{document}

\begin{titlepage}
\noindent
DESY 08-027\\
TTP08-14\\
SFB/CPP-08-21\\
April 2008 \\
\vspace{1.3cm}
\begin{center}
\Large{\bf 
Theoretical status and prospects \\[1ex] 
for top-quark pair production at hadron colliders 
}\\
\vspace{1.5cm}
\large
S. Moch$^{\, a}$ and P. Uwer$^{\, b}$\\
\vspace{1.2cm}
\normalsize
{\it $^a$Deutsches Elektronensynchrotron DESY \\
\vspace{0.1cm}
Platanenallee 6, D--15738 Zeuthen, Germany}\\
\vspace{0.5cm}
{\it $^b$Institut f\"ur Theoretische Teilchenphysik, Universit\"at Karlsruhe\\
\vspace{0.1cm}
D--76128 Karlsruhe, Germany}\\
\vspace{1.8cm}

\large
{\bf Abstract}
\vspace{-0.2cm}
\end{center}
We present an update of the theoretical predictions for the cross section 
of top-quark pair production at Tevatron and LHC.
In particular we employ improvements due to soft gluon resummation 
at next-to-next-to-leading logarithmic accuracy. 
We expand the resummed results and derive analytical finite-order cross sections 
through next-to-next-to-leading order which are exact 
in all logarithmically enhanced terms near threshold.
These results are the best present estimates for the top-quark pair 
production cross section. 
We investigate the scale dependence as well as the sensitivity on the parton luminosities.
\vfill
\end{titlepage}

\newpage

\section{Introduction}
\label{sec:intro}

Top-quark pair production at the LHC is important as the collider 
will accumulate very high statistics for this process. 
In the initial low luminosity run ($\sim 10/(\rm fb\, year)$)
approximately $8 \cdot 10^6$ top-quark pairs will be produced per 
year~\cite{atlas:1999tdr2,cms:2006tdr}.
This data will allow for numerous measurements, e.g. of the top-quark
mass, the electric charge of the top-quark or the weak couplings. 
Furthermore the data will allow precise tests of the production and 
the subsequent decay mechanism including anomalous couplings and top-quark spin correlations.
See for example Ref.~\cite{Quadt:2006jk} for a recent review on
top-quark physics at hadron colliders.

A necessary prerequisite which all these studies share in common is, of course, 
a detailed understanding of the production process. 
In Quantum Chromodynamics (QCD) this includes the radiative corrections 
to the cross section of heavy-quark hadro-production 
at the next-to-leading order (NLO)~\cite{Nason:1988xz,Beenakker:1989bq,Bernreuther:2004jv} 
together with its scale dependence as well as the dependence on the parton
luminosities through the parton distribution functions (PDFs) of the proton. 
Further improvements of the perturbative stability through resummation 
of large Sudakov logarithms to next-to-leading logarithmic (NLL) accuracy 
have been considered as well~\cite{Kidonakis:1997gm,Bonciani:1998vc} 
and employed to generate approximate results 
at the next-to-next-to-leading order (NNLO) in QCD~\cite{Kidonakis:2001nj}.
Very recently also an estimate of bound state effects has been
presented \cite{Hagiwara:2008xy}.

In particular our knowledge on the parton luminosities has constantly improved over the last years. 
Thus, it is an imminent question how these improvements in the 
determination of the parton distribution functions from global fits
affect predictions for physical cross sections at LHC, 
which are sensitive to the gluon distribution function in the regime
where $x\approx 2.5 \cdot 10^{-2}$.
For LHC observables this aspect has been quantitatively approached 
very recently by investigating correlations of rates for top-quark pair production with 
other cross sections~\cite{Nadolsky:2008zw}.
For Tevatron, which to date has provided us with a lot of information 
on the top-quark, most prominently a very precise determination~\cite{tevewwg:2008nq} 
of its mass, $\mt= 172.6 \pm 0.8~(stat.) \pm 1.1~(syst.)$~GeV,
the cross section $\sigma_{pp \to {t\bar t}X}$ in Eq.~(\ref{eq:totalcrs}) 
had been studied some time ago~\cite{Kidonakis:2003qe,Cacciari:2003fi}.
However, due to changes in the available PDF sets from global fits an 
update also seems to be in order here as well.

It is the aim of this article to review theoretical predictions 
for the production cross sections of top-quark pairs at Tevatron and LHC 
and to establish the present theoretical uncertainty.
We provide an update of the NLL resummed cross section as defined 
in Ref.~\cite{Bonciani:1998vc} (and also used in Ref.~\cite{Cacciari:2003fi}). 
Subsequently, we extend these results to the next-to-next-to-leading logarithmic (NNLL) accuracy 
and derive approximate NNLO cross sections thereby improving previous calculations~\cite{Kidonakis:2001nj,Kidonakis:2003qe}.
At two loops we are thus in a position to present all logarithmically enhanced
terms near threshold and to assess their phenomenological impact 
by studying the quality of the perturbative expansion, i.e. the properties of
apparent convergence and the stability under scale variations.
This seems particularly interesting considering not only the anticipated 
experimental precision at LHC~\cite{atlas:1999tdr2,cms:2006tdr} 
but also in view of recent activities aiming at complete NNLO QCD predictions 
for heavy-quark hadro-production~\cite{Dittmaier:2007wz,Czakon:2007ej,Czakon:2007wk,Korner:2008bn,Czakon:2008zk}.

The paper is organized as follows. In Sec.~\ref{sec:status} we set the stage 
and study the threshold sensitivity of the inclusive hadronic cross section 
for top-quark pair production. 
Subsequently, we provide updates of Refs.~\cite{Bonciani:1998vc,Cacciari:2003fi} employing recent sets of PDFs.
In Sec.~\ref{sec:qcdatnnlo} we extend the resummed cross section to NNLL accuracy 
and calculate the complete logarithmic dependence
of the cross section near threshold (including the Coulomb corrections).
Together with the exact NNLO scale dependence of Ref.~\cite{Kidonakis:2001nj} 
these results are the best present estimates for the hadro-production
cross section of top-quark pairs.
We conclude in Sec.~\ref{sec:conclusions} and give some relevant
formulae to the Appendix. In addition cross sections predictions
using different approximations for individual PDFs are also listed
in the Appendix.

\section{Theory status}
\label{sec:status}

Throughout this article, we restrict ourselves to the 
inclusive hadronic cross section $\sigma_{pp \to {t\bar t X}}$ 
(see e.g. Ref.~\cite{Frederix:2007gi} for recent work on top-quark pair 
invariant mass distributions).
Denoting the hadronic center-of-mass energy squared by $\shad $ and
the top-quark mass by $\mt$ the total hadronic cross section for
top-quark pair production is obtained through 
\begin{eqnarray}
  \label{eq:totalcrs}
  \sigma_{pp \to {t\bar t X}}(\shad ,\mt^2) &=& 
  \sum\limits_{i,j = q,{\bar{q}},g} \,\,\,
  \int\limits_{4\mt^2}^{\shad }\,
  d {\shat} \,\, L_{ij}(\shat, \shad, \mufs)\,\,
  \hat{\sigma}_{ij \to {t\bar t}} ({\shat},\mt^2,\mufs,\murs)\, .
\end{eqnarray}
The parton luminosities $L_{ij}(\shat, \shad,  \mufs)$ are defined through
\begin{equation}
  L_{ij}(\shat,\shad,\mufs) = 
  {1\over \shad} \int\limits_{\shat}^\shad 
    {ds\over s} f_{i/p}\left(\mufs,{s\over \shad}\right) 
    f_{j/p}\left(\mufs,{\shat\over s}\right)\, , 
    \label{eq:partonlumi}
\end{equation}
where $f_{i/p}(x,\mufs)$ is the PDF describing the density of partons of flavor $i$ in the 
proton $p$ carrying a fraction $x$ of the initial proton momentum, at
factorization scale $\muf$. Note that we have included $\shad$ into the
definition of $L_{ij}$ to allow an easy comparison of
the luminosity function between different colliders, in particular LHC and Tevatron.
The sum in Eq.~(\ref{eq:totalcrs}) runs over all massless parton flavors and
the top-quark mass used is the so-called pole mass.

Within the context of perturbative QCD the standard way to estimate the 
theoretical uncertainty for the inclusive hadronic cross section 
$\sigma_{pp \to {t\bar t}X}$ in Eq.~(\ref{eq:totalcrs}) 
is based on the residual dependence on the factorization/renormalization scale 
$\muf/\mur$.
Starting from the available predictions to a certain order in
perturbation theory it is common practice to identify the
factorization scale with the renormalization scale
(i.e. $\muf=\mur\equiv\mu$) and to estimate the effect of uncalculated
higher orders by varying $\mu$ in the interval $[\mt/2,2\mt]$.
For a given global PDF fit in contrast, the uncertainties which stem
from uncertainties of the experimental data used in the fits are
treated systematically by a family of $n_{PDF}$ pairs of PDFs, 
where $n_{PDF}$ is the number of parameters used in the fit.
Then, the systematic uncertainty for the observable ${\cal O}$ under 
consideration is estimated by (e.g.~\cite{Pumplin:2002vw,Nadolsky:2008zw}),
\begin{equation}
  \label{eq:pdferr}
  \Delta {\cal O} = 
  \frac{1}{2} \, \sqrt {\sum_{k=1,n_{PDF}} \, 
    ({\cal O}_{k+} - {\cal O}_{k-} )^2}
  \, .
\end{equation}
Here the observable ${\cal O}_{k\pm}$ are obtained by using the parton
distribution functions $f^{k\pm}_{i/p}$ obtained 
by a ``statistical'' $\pm 1\sigma$-variation of the $k$th fit parameter
after diagonalization of the correlation matrix. (Strictly speaking
the fit parameters are varied by an amount which the authors of the
corresponding PDF set take to be equivalent to a $\pm 1\sigma$-variation.)  
To end up with an estimate of the overall uncertainty the uncertainty
coming from the PDFs has to be combined with the uncertainty due
to uncalculated higher orders. Given that the two uncertainties are
very different from each other: in one case we are faced with the traces of
an experimental uncertainty---in the other case we have the systematic
uncertainty due to missing higher order corrections which clearly do
not follow any statistical law. Adding in quadrature the two
uncertainties therefore seems to be inappropriate and we use a linear combination of the 
uncertainties as a conservative estimate for the total uncertainty of the 
top-quark pair cross-section:
\begin{equation}
  \label{eq:range}
  \sigma(2\mt)-\Delta\sigma_{PDF}(2\mt) 
  \, \le \,
  \sigma 
  \, \le \,
  \sigma(\mt/2)+\Delta\sigma_{PDF}(\mt/2)
  \, ,
\end{equation}
where $\Delta\sigma_{PDF}$ is computed according to Eq.~(\ref{eq:pdferr}). 
The NLO QCD corrections for the partonic cross sections $\hat{\sigma}_{ij \to {t\bar t}}$ in Eq.~(\ref{eq:totalcrs}) 
(known since long time~\cite{Nason:1988xz,Beenakker:1989bq,Bernreuther:2004jv}) 
provide the first instance in this procedure where 
a meaningful error can be defined through Eq.~(\ref{eq:range}).

\begin{figure}[htbp]
  \begin{center}
    \includegraphics[width=11.cm]{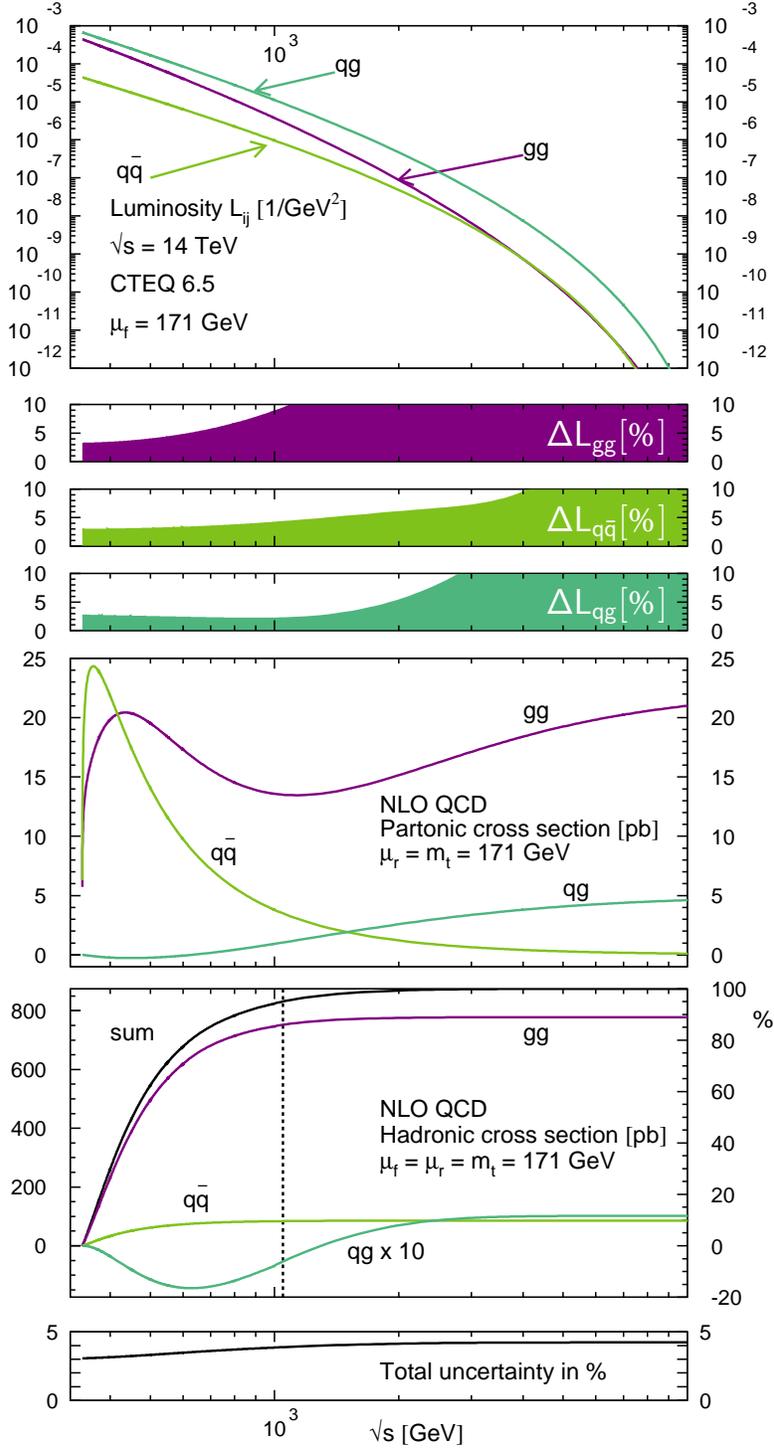}
    \vspace*{-7mm}
    \caption{ \small
      \label{pic:LHCpdf}
      The parton luminosity $L_{ij}$ with the individual PDF uncertainties 
      (upper plots) and the parton cross sections 
      $\hat{\sigma}_{ij \to {t\bar t}}$ 
      at NLO in QCD (third plot from below) as a function of the
      parton energy $\sqrt{s}$.
      The lower plots scan the total cross section 
      $\sigma (\shad ,\mt^2;s_{\max})$ (the total PDF uncertainty) 
      as a function of $\sqrt{s_{\max}}$ for LHC.
      We use $\sqrt{\shad }=14$~TeV,  $\mt=171$~GeV, $\mu = \mt$ and the 
      CTEQ6.5 PDF set.
      The dashed line indicates the value of $\sqrt{s_{\max}}$ for 
      which the cross section is saturated to 95$\%$.
      }
  \end{center}
\end{figure}
\begin{figure}[htbp]
  \begin{center}
    \includegraphics[width=11.cm]{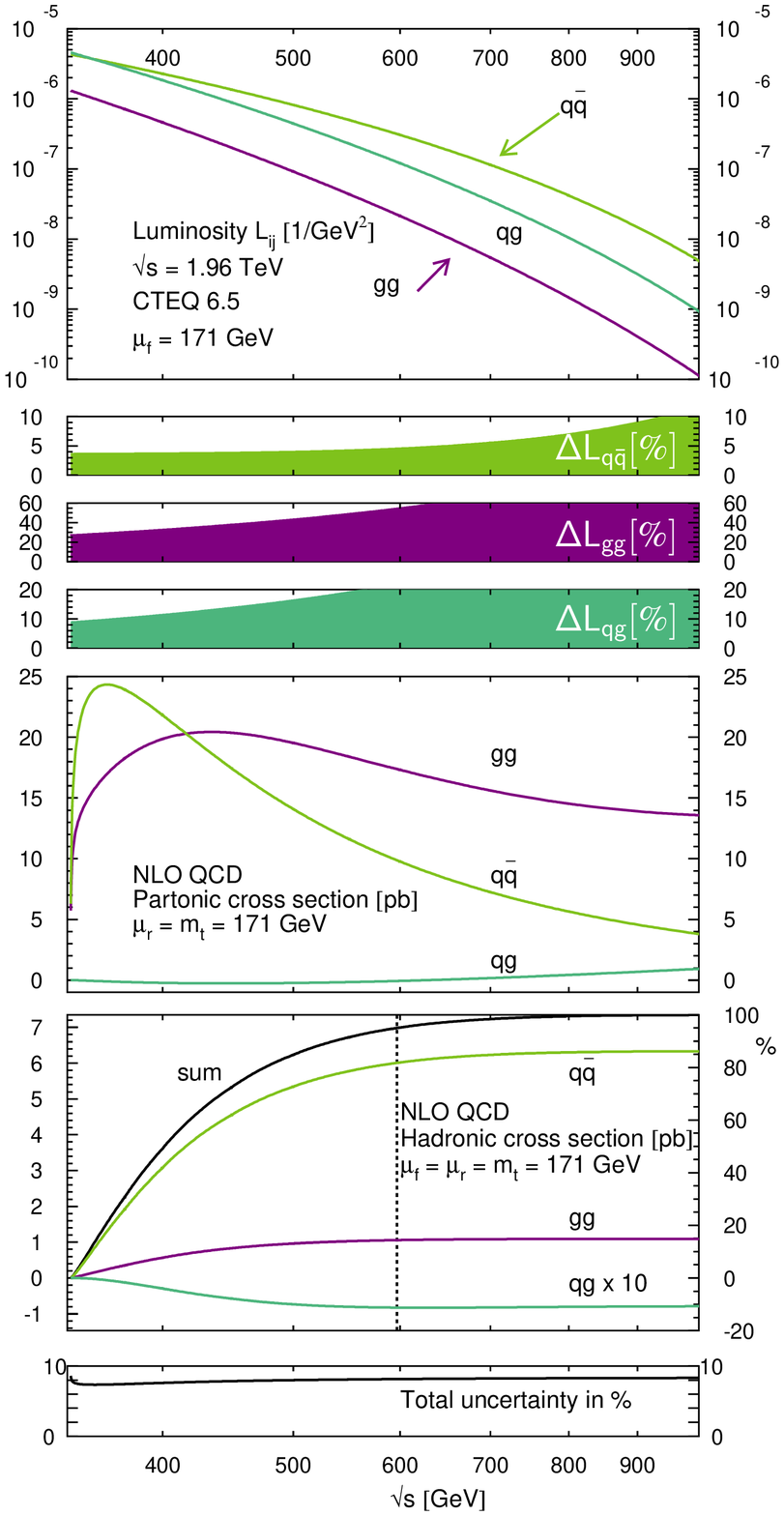}
    \vspace*{-5mm}
    \caption{ \small
      \label{pic:TEVpdf}
      Same as Fig.~\ref{pic:LHCpdf} for Tevatron ($\sqrt{\shad }=1.96$~TeV).
      }
  \end{center}
\end{figure}
Thus, let us start our discussion of the various theoretical uncertainties 
by reviewing at NLO in QCD some basic aspects concerning the parton luminosities
$L_{ij}$ as defined in Eq.~(\ref{eq:partonlumi}) 
at LHC (Fig.~\ref{pic:LHCpdf}) and Tevatron (Fig.~\ref{pic:TEVpdf}). 
At LHC the highest flux is provided by the quark-gluon initial state 
(first plot of Fig.~\ref{pic:LHCpdf}). 
However, as can be seen from the cross section plot of Fig.~\ref{pic:LHCpdf}, 
the parton level cross section $\hat{\sigma}_{{qg} \to {t\bar t}}$ 
is much smaller than for the $q{\bar q}$- or $gg$-initiated processes, 
since the $qg$-channel is of order $\alphas^3$ and thus formally a NLO correction. 
As a consequence, in the total hadronic cross section the $qg$-channel
gives only a contribution at the percent level 
(see second last plot in Fig.~\ref{pic:LHCpdf}). 
In principle the same argument applies for ${\bar q}g$-channel, 
however its contribution to the hadronic cross section is even further 
suppressed because of the smaller parton luminosity.

The second largest parton flux is delivered by the $gg$-channel. 
In combination with the large partonic cross section 
$\hat{\sigma}_{gg \to {t\bar t}}$  this is the most important channel at LHC, 
resulting in about 90\% of all top-quark pairs produced via gluon fusion. 
Close to threshold the uncertainty of the gluon flux 
(estimated using Eq.~\ref{eq:pdferr}, see also discussion there) is about 3\% and at 
$1$~TeV it grows to almost 10\%. 
However, given that the $gg$-channel is largely saturated at parton energies ${\sqrt{\shat}} \simeq 1$~TeV, 
the large PDF uncertainty above $1$~TeV does 
not have significant impact on the overall uncertainty of $\sigma_{pp \to t{\bar t}X}$ in Eq.~(\ref{eq:totalcrs}),
see last plot in Fig.~\ref{pic:LHCpdf}.

The parton luminosity  $L_{q\bar q}$ ranks third at LHC. 
Close to threshold the $q\bar q$-flux is suppressed by roughly a factor 10 compared to $L_{gg}$.
The corresponding parton cross section $\hat{\sigma}_{{q \bar q}\to {t\bar t}}$ vanishes in the high energy limit 
in contrast to the $gg$- and $qg$-case where the partonic cross sections approach a constant at high energies.
Thus, the $q\bar q$-contribution to the hadronic cross section saturates well below $1$~TeV 
and adds the known 10\% at LHC. 
Due to the small numerical contribution of the $q\bar q$- and $qg$-channels the PDF uncertainty of the 
hadronic cross section $\sigma_{pp \to t{\bar t}X}$ is entirely dominated by the uncertainty of $L_{gg}$.

At Tevatron, the situation is reversed (see Fig.~\ref{pic:TEVpdf}). 
The luminosities $L_{ij}$ are ordered in magnitude according to 
$L_{q \bar q} > L_{qg} > L_{gg}$. 
This makes the $q \bar q$-channel by far the dominant one contributing 85\% 
to the hadronic cross section $\sigma_{pp \to t{\bar t}X}$, while gluon fusion almost makes up for the rest.
Although the PDF uncertainty of the $q \bar q$-flux are only 3--4\% at low energies the overall PDF uncertainty of the
top-quark cross section $\sigma_{pp \to t{\bar t}X}$ is large, 
because of the sensitivity to the gluon PDF content at large-$x$, 
which is still poorly constrained at present (see Fig.~\ref{pic:TEVpdf}).

Finally, it is interesting to determine the value of $s_{\max}$ for which 
the cross section 
\begin{eqnarray}
  \sigma(\shad ,\mt^2;s_{\max}) &=& 
  \sum\limits_{i,j = q,{\bar{q}},g} \,\,\,
  \int\limits_{4\mt^2}^{s_{\max }}\,
  d {\shat} \,\, L_{ij}(\shat, \shad, \mufs)\,\,
  \hat{\sigma}_{ij \to {t\bar t}} ({\shat},\mt^2,\mufs,\murs)
\end{eqnarray}
saturates the total cross section 
$\sigma(\shad ,\mt^2)$ to $95\%$.
At Tevatron this happens at ${\sqrt{s_{\max}}} \approx 600$~GeV as 
can be seen from Fig.~\ref{pic:TEVpdf}.
Thus, the total cross section is largely dominated by parton kinematics 
in the range ${\sqrt \shat} \approx 2 \mt$ close to the threshold of the 
top-quark pair.
This makes top-quark pair production at Tevatron an ideal place to apply threshold resummation.
At LHC energies in contrast, the available phase space is larger and 
saturation to $95\%$ 
is only reached at parton energies $\sqrt{s_{\max}} \approx 1$~TeV 
(see Fig.~\ref{pic:LHCpdf}).
This makes the cross section less sensitive to Sudakov logarithms, 
although numerically a significant part still originates 
from the threshold region for parton kinematics 
due to the steeply decreasing parton fluxes.

\medskip

Further refinements of perturbative predictions for $\sigma_{pp \to {t\bar t}X}$ in Eq.~(\ref{eq:totalcrs}) 
do rely on subsequent higher orders to be calculated. 
In particular, the knowledge about large logarithmic corrections 
from regions of phase space near partonic threshold 
allows for improvements of the theoretical accuracy beyond NLO in QCD. 
These Sudakov type corrections can be organized to all orders by means of a 
threshold resummation (e.g. to NLL accuracy~\cite{Kidonakis:1997gm,Bonciani:1998vc}), 
which has been the basis for phenomenological predictions 
employing a resummed cross section as defined in Ref.~\cite{Bonciani:1998vc} (and also used in Ref.~\cite{Cacciari:2003fi}).

However, before updating the NLL resummed results of 
Ref.~\cite{Bonciani:1998vc} 
let us briefly give some relevant resummation formulae.
It is well known that soft gluon resummation for $t {\bar t}$-production relies on 
a decomposition of the parton-level total cross-section in the color basis, 
conveniently defined by color-singlet and color-octet final states. 
Then we can decompose 
\begin{equation}
  \label{eq:singlet-octet}
  \hat{\sigma}_{ij \to {t\bar t}} ({\shat},\mt^2,\mufs,\murs) = 
  \sum\limits_{I=1,8}\, \hat{\sigma}_{ij,\, I}({\shat},\mt^2,\mufs,\murs)\, .
\end{equation}
Moreover, we use the standard definition of Mellin moments 
\begin{eqnarray}
  \label{eq:mellindef}
  \hat{\sigma}^N_{ij,\, I}(\mt^2,\mufs,\murs) &=&
  \int\limits_{0}^{1}\,d\rho\, \rho^{N-1}\,
  \hat{\sigma}_{ij, \,I}(\rho,\mt^2,\mufs,\murs)\, ,
\end{eqnarray}
with 
\begin{equation}
  \rho = {4\mt^2\over \shat}.  
\end{equation}
Then, the resummed Mellin-space cross sections (defined in the \MSbar-scheme) 
for the individual color structures of the scattering process 
are given by a single exponential (see e.g. Refs.~\cite{Contopanagos:1997nh,Catani:1996yz}),
\begin{equation}
\label{eq:sigmaNres}
{\hat{\sigma}^N_{ij,\, I}(\mt^2,\mufs,\murs) \over \hat{\sigma}^{(0), N}_{ij,\, I}(\mt^2,\mufs,\murs)} = 
  g^0_{ij,\, I}(\mt^2,\mufs,\murs) \cdot \exp\, \left( G^{N+1}_{ij,\, I}(\mt^2,\mufs,\murs) \right) + 
  {\cal O}(N^{-1}\ln^n N) \, ,
\end{equation}
where $\hat{\sigma}^{(0), N}_{ij,\, I}$ denotes the Born term and the 
exponents $G^N_{ij,\, I}$ are commonly expressed as 
\begin{equation}
\label{eq:GNexp}
  G^N_{ij\, I}  = 
  \ln N \cdot g^1_{ij}(\lambda)  +  g^2_{ij,\, I}(\lambda)  + 
  a_s\, g^3_{ij,\, I}(\lambda)  + \dots\, ,
\end{equation}
where $\lambda = \beta_0\, a_s\, \ln N$ and $a_s = \alpha_s/(4 \pi)$. 
To NLL accuracy the (universal) functions $g^1_{ij}$ as well as the functions $g^{2}_{ij,\, I}$ 
are relevant in Eq.~(\ref{eq:GNexp}), of course, together with the appropriate 
matching functions $g^0_{ij,\, I}$ in Eq.~(\ref{eq:sigmaNres}).
Explicit expressions can be found below.

For phenomenological applications~\cite{Bonciani:1998vc} the soft-gluon resummation in $N$-space at the parton level 
one introduces an improved (resummed) cross section $\sigmaRES$, 
which is obtained by an inverse Mellin transformation as follows,
\begin{eqnarray}
\label{eq:defsigmares}
\sigmaRES_{ij\to {t\bar t}}(\shat,\mt^2,\mufs,\murs) &=& 
\int\limits_{c-{\rm i}\infty}^{c+{\rm i}\infty}\,
{dN \over 2\pi {\rm i}}\, \rho^{-N+1}\,
  \sum\limits_{I=1,8}\, \left(
  \hat{\sigma}^N_{ij,\, I}(\mt^2,\mufs,\murs) - 
  \hat{\sigma}^N_{ij,\, I}(\mt^2,\mufs,\murs)\biggr|_{\rm NLO}
  \right) 
\nonumber\\ &&
  + 
  \hat{\sigma}^{\rm NLO}_{ij \to {t\bar t}}({\shat},\mt^2,\mufs,\murs)\, ,
\end{eqnarray}
where $\hat{\sigma}^{\rm NLO}_{ij \to {t\bar t}}$ is the standard fixed order 
cross section at NLO in QCD and $\hat{\sigma}^N_{ij,\, I}\bigr|_{\rm NLO}$ is 
the perturbative truncation of Eq.~(\ref{eq:sigmaNres}) at the same order in $\alpha_s$.
Thus, the right-hand side of Eq.~(\ref{eq:defsigmares}) reproduces the
fixed order results and resums soft-gluon effects beyond NLO to NLL accuracy.

\begin{figure}[htb]
  \begin{center}
    \includegraphics[width=0.49\textwidth]{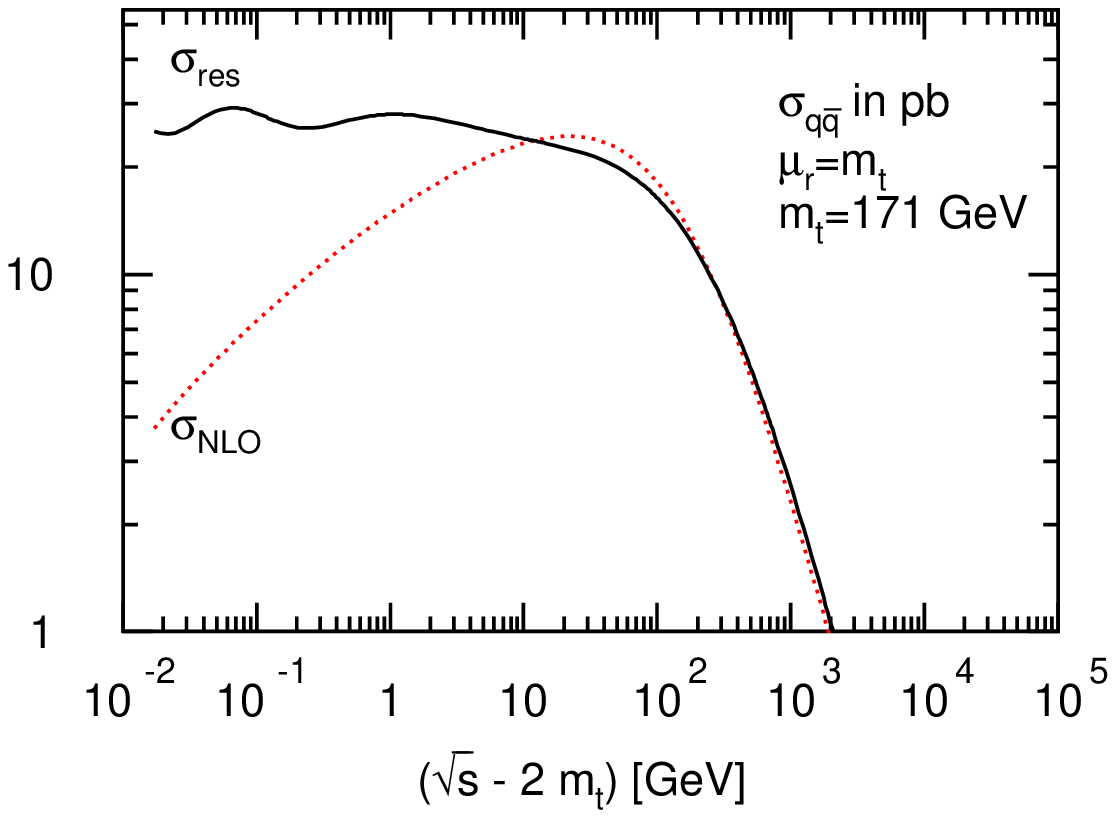}
    \includegraphics[width=0.49\textwidth]{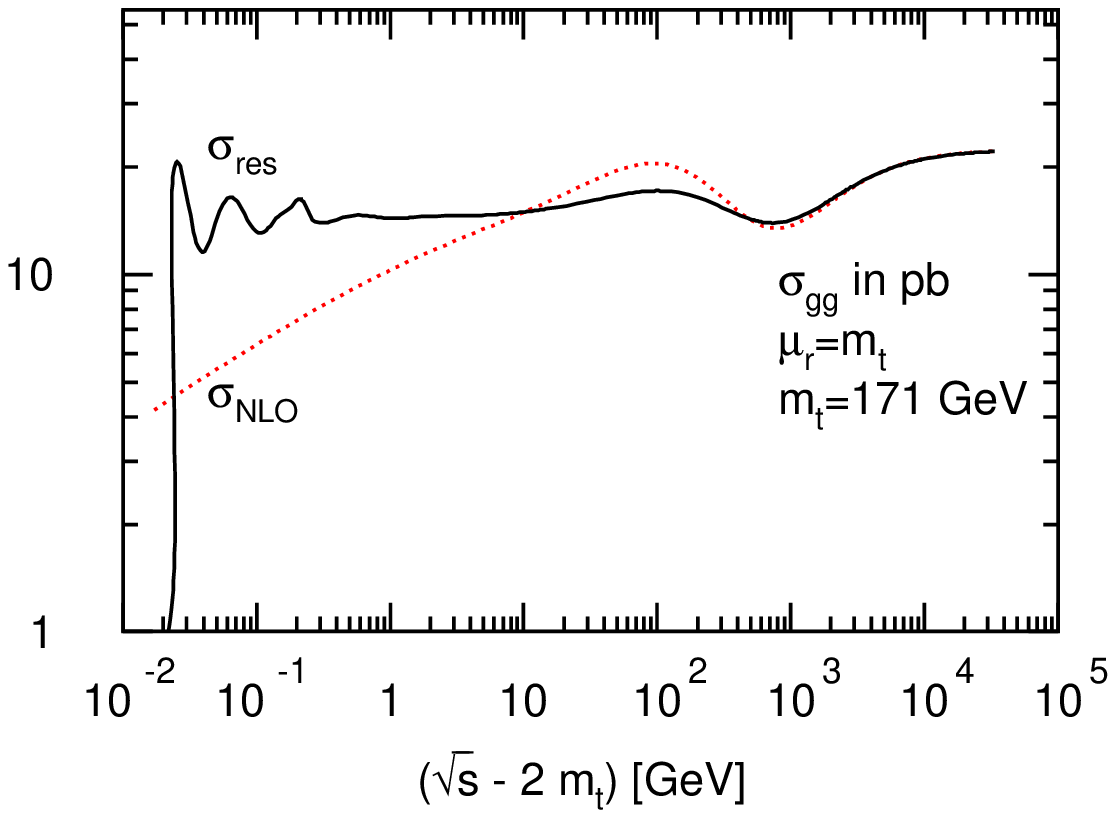}
    \vspace*{-1mm}
    \caption{ \small
      \label{pic:r-partonic-crs}
      The partonic cross sections for the processes 
      ${q\bar q} \to {t\bar t}$ (left) and $gg \to {t\bar t}$ (right)
      in pb, and for $\mu = \mt=171$~GeV. 
      The dotted lines are the exact NLO result~\cite{Nason:1988xz,Beenakker:1989bq} and the solid lines 
      correspond to the NLL resummed cross sections~\cite{Bonciani:1998vc} 
      (see text for details).
      }
    \vspace*{2mm}
  \end{center}
\end{figure}
In Fig.~\ref{pic:r-partonic-crs} we plot the resummed cross section 
$\sigmaRES_{ij\to {t\bar t}}$ 
as defined in Eq.~(\ref{eq:defsigmares}) zooming in on the threshold region.
We display the ${q\bar q}$-channel (left) and the $gg$-channel (right) as a function of 
the distance from the partonic threshold at $\sqrt{\shat} = 2 \mt$. 
As mentioned already the $qg$-channel is suppressed by an additional power of $\alphas$ 
and therefore does not contribute large Sudakov logarithms to the accuracy considered here.

\medskip

The results for $\sigmaRES$ as shown in Fig.~\ref{pic:r-partonic-crs}
have been obtained by performing the inverse Mellin transform in
Eq.~(\ref{eq:defsigmares}) numerically. 
Following the procedure described in Ref.~\cite{Bonciani:1998vc} 
we have compared our results obtained from the numerical inversion with the
ones shown in Ref.~\cite{Bonciani:1998vc} and found complete agreement.
To be precise, the treatment of the constant terms in 
Ref.~\cite{Bonciani:1998vc} 
(i.e. the terms denoted $g^0_{ij,\, I}$ in Eq.~(\ref{eq:sigmaNres})) 
differs slightly from the minimal approach, e.g. Eq.~(\ref{eq:defsigmares}).
Some constants which are formally subleading have been included in 
Ref.~\cite{Bonciani:1998vc},
and moreover, several schemes for power suppressed terms in $N$ have been implemented.
The resummed result shown in Fig.~\ref{pic:r-partonic-crs} is defined through
Eq.~(63) of Ref.~\cite{Bonciani:1998vc} with the parameter $A$ set to 2.
Another issue concerns the precise numerical matching of the exact NLO cross section and
the resummed result in Eq.~(\ref{eq:defsigmares}). 
We apply 
the resummed result only for $\sqrt{\shat} - 2 \mt \le 10$ 
GeV. The exact point is determined from the crossing of the two
curves for \sigmaRES\ and \sigmaNLO. Note that the precise numerical
value is not important. 
Effectively $\sigmaRES$ is thereby restricted to a region of parton
energies of $\sqrt{\shat}\approx 2\mt$ 
or in other words, to a kinetic energy of one top-quark of a 
few GeV.

\begin{table}[htbp]
  \begin{center}
    \leavevmode
    \begin{tabular}{c|llc|llc|llc} \hline
 &\multicolumn{3}{|c|}{only scale uncertainty}&\multicolumn{3}{|c|}{only pdf uncertainty}&\multicolumn{3}{|c}{total uncertainty}\\ 
m &min&max&$\delta [\%]$&min&max&$\delta [\%]$&min&max&$\delta [\%]$\\ \hline\hline
165&8.29&9.41&7&8.58&9.61&6&7.86&9.99&12\\
166&8.03&9.11&7&8.31&9.3&6&7.61&9.67&12\\
167&7.78&8.83&7&8.06&9.01&6&7.38&9.37&12\\
168&7.54&8.55&7&7.81&8.73&6&7.15&9.07&12\\
169&7.31&8.28&7&7.57&8.46&6&6.93&8.79&12\\
170&7.09&8.03&7&7.34&8.2&6&6.72&8.51&12\\
171&6.87&7.78&7&7.12&7.94&6&6.52&8.25&12\\
172&6.66&7.54&7&6.9&7.7&6&6.32&7.99&12\\
173&6.46&7.31&7&6.69&7.46&6&6.13&7.75&12\\
174&6.26&7.09&7&6.49&7.23&6&5.94&7.51&12\\
175&6.07&6.87&7&6.3&7.01&6&5.77&7.28&12\\
176&5.89&6.67&7&6.11&6.8&6&5.59&7.06&12\\
177&5.71&6.47&7&5.93&6.59&6&5.43&6.84&12\\
178&5.54&6.27&7&5.75&6.4&6&5.26&6.64&12\\
179&5.38&6.08&7&5.58&6.2&6&5.11&6.44&12\\
180&5.22&5.9&7&5.41&6.02&6&4.96&6.24&12\\
\hline
\end{tabular}

  \end{center}
  \caption{ \small
      \label{tab:CTEQ6.5-MATCHED-Tevatron}
      The NLL resummed cross section of Ref.~\cite{Bonciani:1998vc} 
      in pb (see text for details) 
      for various values of the top-quark mass $\mt$ at Tevatron
      ($\sqrt{ \shad}=1.96$~TeV) using the CTEQ6.5 PDF set~\cite{Tung:2006tb}.
      $\delta$ is the relative uncertainty with respect to the central
      value: $\delta = 100\times 
      (\mbox{max}-\mbox{min})/(\mbox{max}+\mbox{min})$.
      }
\end{table}
\begin{table}[htbp]
  \begin{center}
    \leavevmode
    \begin{tabular}{c|llc|llc|llc} \hline
 &\multicolumn{3}{|c|}{only scale uncertainty}&\multicolumn{3}{|c|}{only pdf uncertainty}&\multicolumn{3}{|c}{total uncertainty}\\ 
m &min&max&$\delta [\%]$&min&max&$\delta [\%]$&min&max&$\delta [\%]$\\ \hline\hline
165&8.53&9.87&8&9.19&9.7&3&8.32&10.1&10\\
166&8.26&9.56&8&8.9&9.38&3&8.05&9.83&10\\
167&8&9.25&8&8.62&9.08&3&7.8&9.51&10\\
168&7.74&8.95&8&8.34&8.8&3&7.55&9.21&10\\
169&7.5&8.67&8&8.08&8.52&3&7.31&8.92&10\\
170&7.26&8.4&8&7.83&8.25&3&7.08&8.63&10\\
171&7.04&8.13&8&7.58&7.99&3&6.86&8.36&10\\
172&6.82&7.87&8&7.35&7.74&3&6.65&8.1&10\\
173&6.6&7.63&8&7.12&7.5&3&6.44&7.84&10\\
174&6.4&7.39&8&6.9&7.26&3&6.24&7.6&10\\
175&6.2&7.16&8&6.69&7.04&3&6.05&7.36&10\\
176&6.01&6.94&8&6.48&6.82&3&5.87&7.13&10\\
177&5.83&6.73&8&6.28&6.61&3&5.69&6.91&10\\
178&5.65&6.52&8&6.09&6.41&3&5.51&6.7&10\\
179&5.48&6.32&8&5.9&6.21&3&5.35&6.49&10\\
180&5.31&6.13&8&5.72&6.02&3&5.18&6.29&10\\
\hline
\end{tabular}

  \end{center}
  \caption{ \small 
    \label{MRST06nnlo-MATCHED-Tevatron}
    Same as in Tab.~\ref{tab:CTEQ6.5-MATCHED-Tevatron} using the MRST-2006 NNLO PDF set~\cite{Martin:2007bv}.
    }
\end{table}
\begin{table}[htbp]
  \begin{center}
    \leavevmode
    \begin{tabular}{c|llc|llc|llc} \hline
 &\multicolumn{3}{|c|}{only scale uncertainty}&\multicolumn{3}{|c|}{only pdf uncertainty}&\multicolumn{3}{|c}{total uncertainty}\\ 
m &min&max&$\delta [\%]$&min&max&$\delta [\%]$&min&max&$\delta [\%]$\\ \hline\hline
165&937&1154&11&1006&1074&4&906&1191&14\\
166&911&1122&11&978&1044&4&881&1159&14\\
167&886&1091&11&951&1016&4&856&1127&14\\
168&862&1061&11&925&988&4&833&1096&14\\
169&838&1032&11&900&962&4&810&1066&14\\
170&816&1004&11&875&936&4&788&1038&14\\
171&794&977&11&852&911&4&767&1010&14\\
172&773&950&11&829&887&4&746&983&14\\
173&752&925&11&806&863&4&726&957&14\\
174&732&900&11&785&841&4&707&931&14\\
175&713&877&11&764&819&4&688&907&14\\
176&694&853&11&744&797&4&670&883&14\\
177&676&831&11&724&777&4&653&860&14\\
178&659&809&11&705&757&4&636&838&14\\
179&642&788&11&687&737&4&619&816&14\\
180&625&768&11&669&718&4&603&795&14\\
\hline
\end{tabular}

  \end{center}
  \caption{ \small
      \label{tab:CTEQ6.5-MATCHED-LHC}
      The NLL resummed cross section of Ref.~\cite{Bonciani:1998vc} in pb (see text for details) 
      for various values of the top-quark mass $\mt$ at LHC 
      ($\sqrt{ \shad}=14$~TeV) using the CTEQ6.5 PDF set~\cite{Tung:2006tb}.
    }
\end{table}
\begin{table}[htbp]
  \begin{center}
    \leavevmode
    \begin{tabular}{c|llc|llc|llc} \hline
 &\multicolumn{3}{|c|}{only scale uncertainty}&\multicolumn{3}{|c|}{only pdf uncertainty}&\multicolumn{3}{|c}{total uncertainty}\\ 
m &min&max&$\delta [\%]$&min&max&$\delta [\%]$&min&max&$\delta [\%]$\\ \hline\hline
165&986&1222&11&1084&1110&2&974&1236&12\\
166&959&1189&11&1054&1080&2&947&1203&12\\
167&933&1156&11&1026&1050&2&921&1170&12\\
168&908&1125&11&998&1022&2&896&1138&12\\
169&883&1094&11&971&995&2&872&1108&12\\
170&860&1065&11&945&968&2&849&1078&12\\
171&837&1036&11&920&943&2&826&1049&12\\
172&815&1009&11&895&918&2&804&1021&12\\
173&793&982&11&872&894&2&783&994&12\\
174&773&956&11&849&870&2&763&968&12\\
175&753&931&11&827&848&2&743&943&12\\
176&733&907&11&805&826&2&723&918&12\\
177&714&883&11&784&805&2&705&895&12\\
178&696&860&11&764&784&2&686&872&12\\
179&678&838&11&744&764&2&669&849&12\\
180&661&817&11&725&745&2&652&828&12\\
\hline
\end{tabular}

  \end{center}
  \caption{ \small
    \label{MRST06nnlo-MATCHED-LHC}
    Same as in Tab.~\ref{tab:CTEQ6.5-MATCHED-LHC} using the MRST-2006 NNLO PDF set~\cite{Martin:2007bv}.
  }
\end{table}
\begin{figure}[htb]
  \begin{center}
    \includegraphics[width=0.49\textwidth]{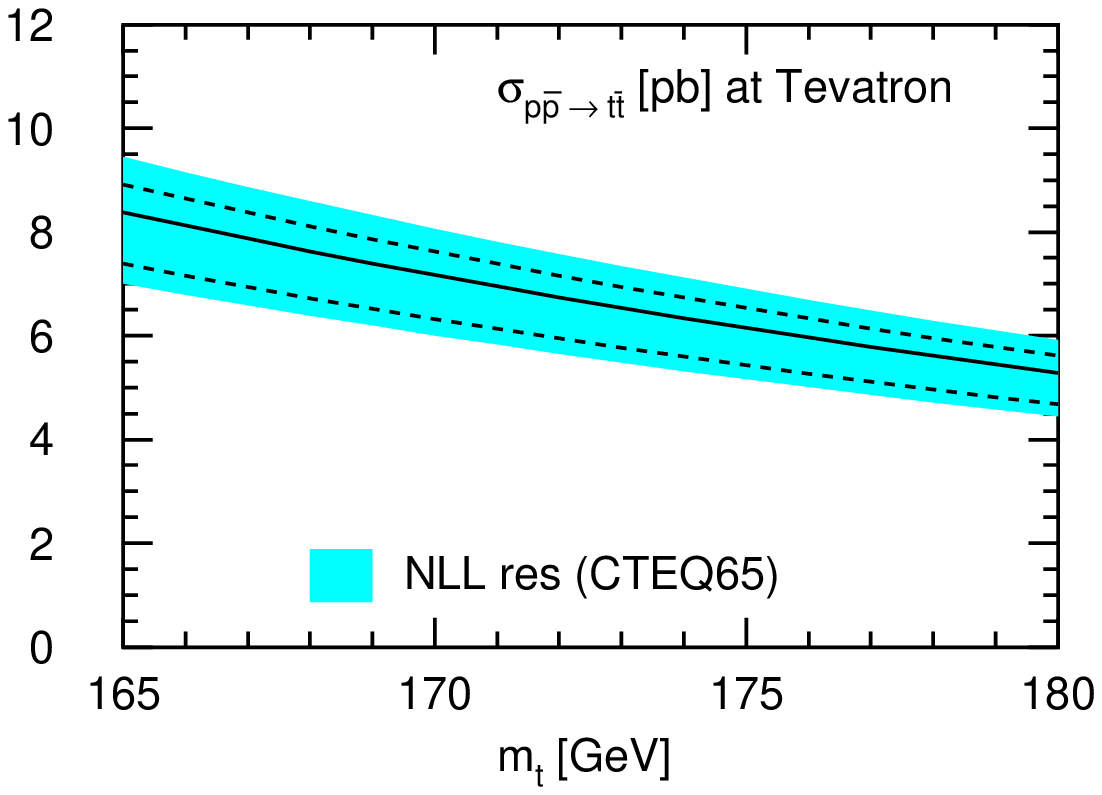}
    \includegraphics[width=0.49\textwidth]{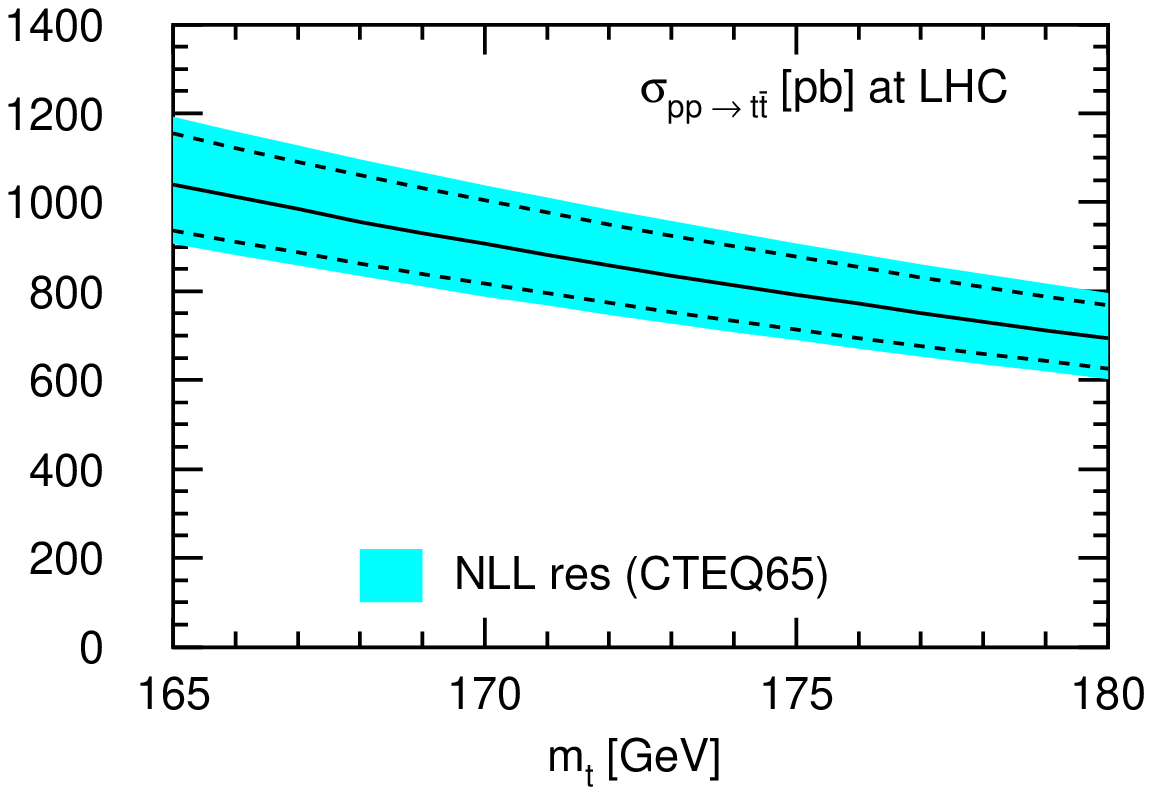}
  \vspace*{-1mm}
  \caption{ \small
    \label{pic:ttbar-resum}
    The $t{\bar t}$ total cross section resummed to NLL 
    accuracy~\cite{Bonciani:1998vc} as a function of $\mt$ for the Tevatron
    at $\sqrt{ \shad}=1.96$~TeV (left) and LHC at $\sqrt{ \shad}=14$~TeV (right).
    The solid line is the central value for $\mu=\mt$, the dashed
    lower and upper lines correspond to $\mu=2\mt$ and $\mu=\mt/2$, 
    respectively. The band denotes the total uncertainty that is the 
    uncertainty due to scale variations and the PDF uncertainty of
    the CTEQ6.5 set~\cite{Tung:2006tb} combined together according to
    Eq.~(\ref{eq:range}).
    }
\vspace*{2mm}
  \end{center}
\end{figure}
It should be stressed, that the total hadronic cross section is not very sensitive 
to these fine details due to the convolution with the parton luminosities $L_{ij}$ in Eq.~(\ref{eq:totalcrs}). 
However, to per mille accuracy (see Tabs.~\ref{tab:CTEQ6.5-MATCHED-Tevatron}--\ref{MRST06nnlo-MATCHED-LHC}) 
these details become noticeable.
Having them clarified, we are now in a position to update previous results~\cite{Bonciani:1998vc,Cacciari:2003fi} 
for the $t {\bar t}$-cross section using modern PDFs, such as the CTEQ6.5 PDF set~\cite{Tung:2006tb}. 
In comparison to older sets, we find e.g. a shift of 3\% in the total cross section 
between the PDF sets CTEQ6.5~\cite{Tung:2006tb} and
CTEQ6.1~\cite{Pumplin:2002vw}, see also Ref.~\cite{Nadolsky:2008zw}.
Since Eq.~(\ref{eq:defsigmares}) effectively contains the dominant part of higher orders (NNLO and beyond) 
it seems however equally appropriate to use also the MRST-2006 NNLO PDF set~\cite{Martin:2007bv}.
From the results in
Tabs.~\ref{tab:CTEQ6.5-MATCHED-Tevatron}--\ref{MRST06nnlo-MATCHED-LHC} and Fig.~\ref{pic:ttbar-resum} 
we conclude that the present overall uncertainty on the NLL resummed $t {\bar t}$-cross section is 12\% at Tevatron 
with a small shift of the central value between different PDF sets.
At LHC the NLL resummed cross section of Ref.~\cite{Bonciani:1998vc} reduces
effectively to the NLO QCD prediction of Refs.~\cite{Nason:1988xz,Beenakker:1989bq} with an overall uncertainty of 14\%. 
This is due to the small gluon contribution close to threshold (see Fig.~\ref{pic:r-partonic-crs}) 
and the cut when matching NLO and the $\sigmaRES$ in the numerical determinations. 
Thus the reduction of the theoretical uncertainty through Eq.~(\ref{eq:defsigmares}) is marginal in this case. 
We will see in the next Section~\ref{sec:qcdatnnlo}, how this situation can be improved with the
help of approximate NNLO QCD corrections.

\section{Prospects at NNLO in QCD}
\label{sec:qcdatnnlo}

Let us now extend the theory predictions for heavy-quark hadro-production. 
We will focus on the threshold region and improve soft gluon resummation to NNLL accuracy. 
Subsequently, we employ the resummed cross section to generate higher order
perturbative corrections -- more specifically an approximate NNLO cross section $\sigmaNNLO$ 
which is exact to logarithmic accuracy (including the Coulomb corrections).
To that end, we briefly recall the steps leading to the final form for $G^N_{ij,\, I}$ 
in Eq.~(\ref{eq:GNexp}).
In order to achieve NNLL accuracy the function $g^3_{ij,\, I}$ is of particular interest here.

The exponential $G^N_{ij,\, I}$ in Eq.~(\ref{eq:sigmaNres}) is build up from universal
radiative factors for the individual color structures which take the form
\begin{eqnarray}
  \label{eq:GNQQbar}
  G^N_{q{\bar q}/gg,\, I} & = & G^N_{\rm DY/Higgs} + \delta_{I,8} G^N_{Q{\bar Q}} \, ,
\end{eqnarray}
where the exponentiation of singlet contribution (i.e. a colorless massive final state) 
follows from the Drell-Yan process and hadronic Higgs production in gluon fusion.
The corresponding functions $G^N_{\rm DY}$ and $G^N_{\rm Higgs}$ are very well-known~\cite{Vogt:2000ci,Catani:2003zt,Moch:2005ba}.
The exponentiation of the color-octet contribution receives an additional 
contribution $G^N_{Q{\bar Q}}$ due to soft gluon emission from the heavy-quark pair in the final state.
The final-state system carries a total color charge given by the $Q{\bar Q}$ charge, 
thus its contribution to soft radiation vanishes in the color-singlet channels
regardless of the initial state partons.
Moreover gluon emission from massive quarks does not lead to collinear logarithms 
and therefore $G^N_{Q{\bar Q}}$ starts at NLL accuracy only.
Explicit formulae are
\begin{eqnarray}
  \label{eq:GNresDYH}
  G^N_{\rm DY/Higgs} &=&
  \int\limits_0^1 \! dz \, {z^{N-1}-1 \over 1-z} \,\int_{\mu_f^{\,2}}^{4\mt^2 (1-z)^2}
  {dq^2 \over q^2}\,   2\, A_{i}(\alpha_s(q^2)) 
  + D_{i}(\alpha_s(4\mt^2[1-z]^2))
  \, ,
\\
  \label{eq:GNresQQbar}
G^N_{Q{\bar Q}} &=&   \int\limits_0^1 \! dz\, {z^{N-1}-1 \over 1-z} \, D_{Q{\bar Q}}(\alpha_s(4\mt^2[1-z]^2))
\end{eqnarray}
with anomalous dimensions $A_{i}$ and $D_{i}$, $i = q,g$ corresponding to DY and Higgs, respectively.
The effects of collinear soft-gluon radiation off initial-state partons $i = q,g$ are collected by
the first term in Eq.~(\ref{eq:GNresDYH}) while the process-dependent contributions from large-angle soft
gluons are resummed by the second term. 
Soft radiation from the heavy-quark pair in the final state is summarized by $G^N_{Q{\bar Q}}$ in Eq.~(\ref{eq:GNresQQbar}) 
with the corresponding anomalous dimension $D_{Q{\bar Q}}$. 

The extension to NNLL requires the $A_q$ and $A_g$ to three loops~\cite{Moch:2004pa,Vogt:2004mw}
and the function $D_q$ and $D_g$ to two loops~\cite{Vogt:2000ci,Catani:2003zt,Moch:2005ba} 
(the latter are actually known to three loops as well~\cite{Moch:2005ky,Laenen:2005uz}).
Explicit expressions using the expansions
\begin{equation}
\label{eq:aexp}
  f(\alpha_s) = 
  \sum\limits_{l}\, f^{(l)}\, {\alpha_s^{l} \over 4\pi} \equiv 
  \sum\limits_{l}\, f^{(l)}\, a_s^{l} \, .
\end{equation}
are collected in Eqs.~(\ref{eq:Aqexp})--(\ref{eq:AgDg}) in the Appendix.
The remaining anomalous dimensions $D_{Q \bar{Q}}$ for soft radiation off a heavy-quark pair in the final state 
is needed to two loops.
For the latter we use the exact calculation of the two-loop QCD corrections
to the massive heavy quark form factor~\cite{Bernreuther:2004ih} along 
with Ref.~\cite{Mitov:2006xs}, where the exponentiation of the form factor for
massive colored particles in the limit $m^2 \to 0$ has been clarified.
From the all-order singularity structure of the massive form factor~\cite{Mitov:2006xs}, 
we can read off the single poles corresponding to soft gluon emission at one and two loops.
After a trivial substitution of color factors ($C_A$ for $C_F$) 
we find 
\begin{eqnarray}
  \label{eq:dqqbar}
  D^{(1)}_{Q{\bar Q}} = - A_g^{(1)}\, ,\qquad\qquad
  D^{(2)}_{Q{\bar Q}} = - A_g^{(2)}\, .
\end{eqnarray}
As pointed out in Ref.~\cite{Bonciani:1998vc} the one-loop value of $D_{Q{\bar Q}}$
agrees with Ref.~\cite{Kidonakis:1997gm} where the soft anomalous dimension matrix (in color space) 
$\Gamma^{(1)}_{IJ}$ for heavy-quark production has been calculated at order $\alpha_s$.
In the limit $\beta \to 0$ the matrix $\Gamma^{(1)}_{IJ}$ diagonalizes in the
singlet-octet basis and reproduces $D^{(1)}_{Q{\bar Q}}$ from its eigenvalue in the octet channel.
Moreover, the structure of $D_{Q{\bar Q}}$ as determined
from Ref.~\cite{Mitov:2006xs} agrees also with the two-loop soft anomalous
dimension matrix $\Gamma^{(2)}_{IJ}$ for vanishing parton masses which obeys the 
following factorization property~\cite{MertAybat:2006wq,MertAybat:2006mz},
\begin{equation}
  \label{eq:masslessGamma2}
\Gamma_{IJ}\biggr|_{m=0} = a_s\, \Gamma^{(1)}_{IJ}\biggr|_{m=0} \left(1 + a_s {A_g^{(2)} \over A_g^{(1)}} \right) \, ,
\end{equation}
with the well-known ratio ${A_g^{(2)}/A_g^{(1)}}$~\cite{Kodaira:1982nh}. 
As a further check on Eq.~(\ref{eq:dqqbar}) it would, of course, be very interesting 
to repeat the calculation of Ref.~\cite{MertAybat:2006wq,MertAybat:2006mz} 
for heavy-quark hadro-production at two loops, i.e. with non-vanishing parton 
masses $m \neq 0$.

\begin{figure}[htb]
  \begin{center}
    \includegraphics[width=0.49\textwidth]{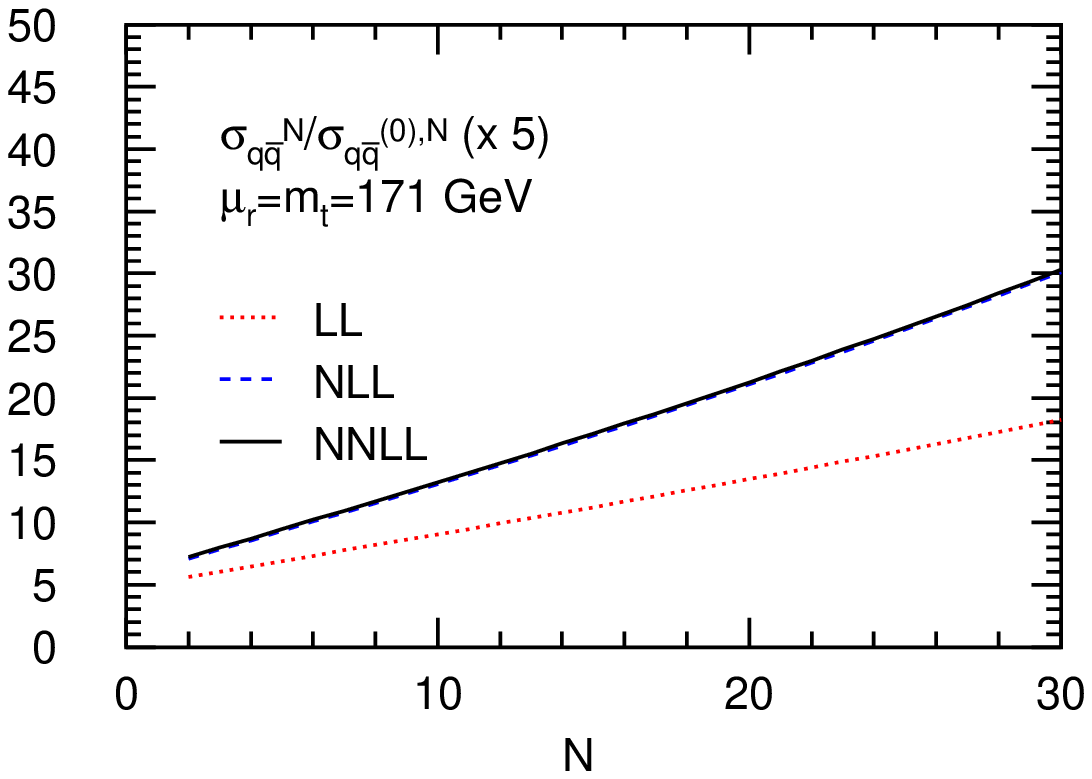}
    \includegraphics[width=0.49\textwidth]{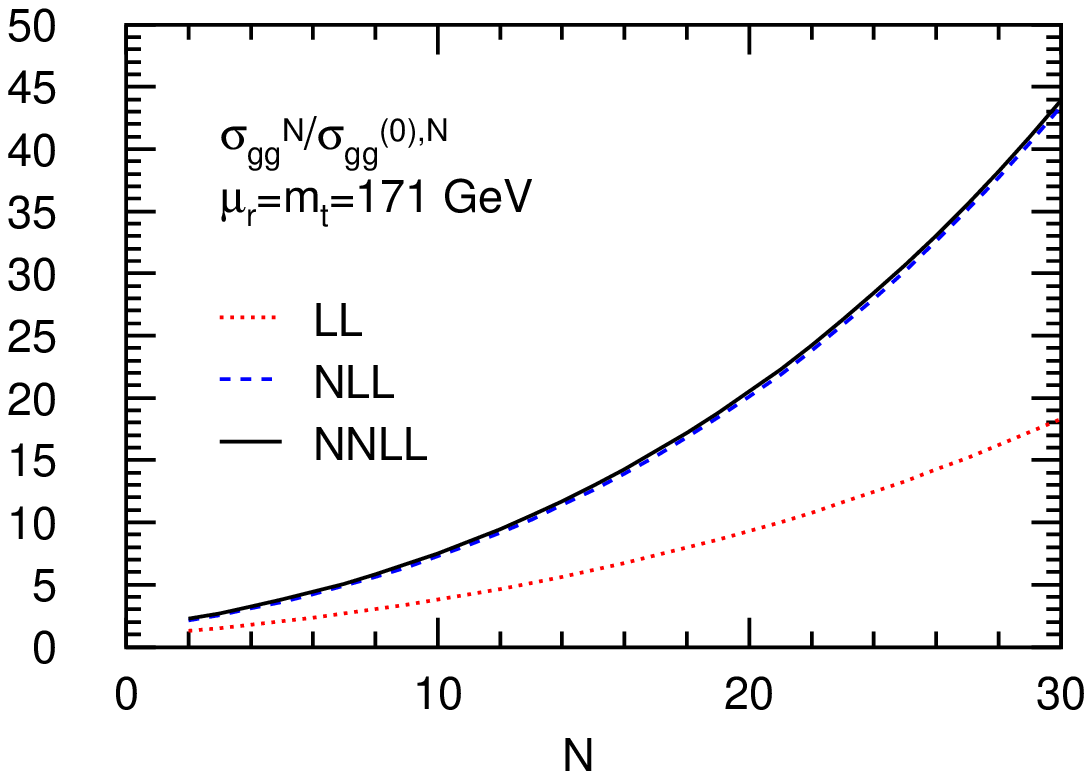}
    \vspace*{-1mm}
    \caption{ \small 
      \label{pic:nnllGN}
      The resummed cross section in Mellin space for the $q{\bar q}$ (left)
      and the $gg$-channel (right) normalized to the Born result 
      for $\mu = \mt=171$~GeV.
      The dotted lines are the LL approximation, the dashed lines denote the
      NLL result and the solid lines correspond 
      to the NNLL result derived in this paper.
      }
    \vspace*{2mm}
  \end{center}
\end{figure}
Finally, explicit integration of Eqs.~(\ref{eq:GNresDYH})--(\ref{eq:GNresQQbar}) leads
to the functions $g^1_{ij}$, $g^2_{ij,\, I}$, $g^3_{ij,\, I}$ of Eq.~(\ref{eq:GNexp}) 
and to the matching $g^0_{ij,\, I}$ in Eq.~(\ref{eq:sigmaNres}) 
which we collected in Eqs.~(\ref{eq:g1res})--(\ref{eq:g0res}) in the Appendix. 
All formulae can be obtained by simple substitutions e.g. from deep-inelastic scattering (DIS) 
in Ref.~\cite{Moch:2005ba}.
In Fig.~\ref{pic:nnllGN} we display the resummed cross section of Eq.~(\ref{eq:sigmaNres}) 
(normalized to the respective Born result) 
for the $q{\bar q}$ and the $gg$-channel in increasing logarithmic accuracy. 
Fig.~\ref{pic:nnllGN} clearly shows the good convergence property of the NNLL contribution 
similar to other observables investigated previously~\cite{Vogt:2000ci,Catani:2003zt,Moch:2005ba}.
As a matter of fact, Eq.~(\ref{eq:GNexp}) may even be extended to next-to-next-to-next-to-leading 
accuracy (N$^3$LL), as the relevant functions $g^4_{ij,\, I}$ can be
easily derived from Ref.~\cite{Moch:2005ba} 
and the respective anomalous dimensions are known.
Following the arguments based on the exponentiation of the form factor 
for massive colored partons~\cite{Mitov:2006xs} leading to Eq.~(\ref{eq:dqqbar}) we identify 
$D^{(3)}_{Q{\bar Q}} = - A_g^{(3)}$ 
and for the four-loop terms $A_q^{(4)}$ and $A_g^{(4)}$ a Pad\'e estimate exists~\cite{Moch:2005ba}.
However, presently we are lacking knowledge on the matching functions $g^0_{ij}$ at this order.
From experience we expect N$^3$LL effects to be numerically very small, though, 
and we leave this issue to future investigations.

\medskip

Let us instead use the resummed cross section $\sigmaRES$ (now known at NNLL accuracy) 
to construct an approximate NNLO cross section $\sigmaNNLO$ by expanding Eq.~(\ref{eq:sigmaNres}) to second order. 
In this way, we determine the corresponding Sudakov logarithms appearing in the NNLO corrections, 
i.e. the powers of $\ln^k N$ in Mellin space 
or $\ln^k \beta$ in momentum space with $k=1,...,4$ and the velocity
of the heavy quark 
\begin{displaymath}
  \beta=\sqrt{1-4\mt^2/s}.  
\end{displaymath}
As a technical remark, we stress here that in a fixed order expansion the inverse Mellin transformation, 
i.e. the mapping of powers of $\ln N$ to powers of $\ln(\beta)$ can be uniquely performed. 
The Mellin-space accuracy up to power suppressed terms in $N$ corresponds to
neglecting higher order polynomials in $(1 - \rho)$ with 
$\rho = 4 \mt^2/\shat$. 
We give some formulae in the Appendix.
Moreover, we can even include the complete Coulomb corrections at two
loops thanks to Ref.~\cite{Bernreuther:2004ih,Czarnecki:1997vz}.
In the singlet-octet decomposition $\hat{\sigma}_{ij,\, I}$ 
of the individual color structures in the cross section, 
the result of Ref.~\cite{Bernreuther:2004ih,Czarnecki:1997vz} can be directly applied to the singlet
case, while a simple modification of the color factors (that is $(C_F-C_A/2)$ instead of $C_F$) 
accounts for the octet case (see e.g. \cite{Pineda:2006ri}).

Thus, we are in a position to present the threshold expansion of the inclusive partonic cross sections 
$\hat{\sigma}_{ij \to {t\bar t}} (s,m^2,\mu^2)$ entering Eq.~(\ref{eq:totalcrs}).
In the perturbative expansions in powers of the strong coupling constant 
$\alphas$ as defined in Eq.~(\ref{eq:aexp}) and setting $\mu = \mt$ and $n_f = 5$, 
we have for the $q\bar{q}$ channel in the \MSbar-scheme
\begin{eqnarray}
\label{eq:sqq10-num}
  \hat{\sigma}^{(1)}_{q\bar{q}\to {t\bar t}} &=& 
  \hat{\sigma}^{(0)}_{q\bar{q}\to {t\bar t}} 
  \Biggr\{ 
         42.667 \* \ln^2 \beta
       - 20.610 \* \ln \beta
       + 13.910 - 3.2899 \* {1 \over \beta}
    \Biggr\}\, ,
\\
\label{eq:sqq20-num}
  \hat{\sigma}^{(2)}_{q\bar{q}\to {t\bar t}} &=& 
  \hat{\sigma}^{(0)}_{q\bar{q}\to {t\bar t}} 
  \Biggr\{ 
         910.22 \* \ln^4 \beta
       - 1315.5 \* \ln^3 \beta
       + \left( 565.80 - 140.37 \* {1 \over \beta} \right) \* \ln^2 \beta
\nonumber \\ &&\qquad\qquad
       + \left( 862.42 + 32.106 \* {1 \over \beta} \right) \* \ln \beta
       + 3.6077 \* {1 \over \beta^2} + 10.474 \* {1 \over \beta} + C^{(2)}_{q{\bar q}}
    \Biggr\}\, ,
\\
\label{eq:sgg10-num}
  \hat{\sigma}^{(1)}_{gg\to {t\bar t}} &=& 
  \hat{\sigma}^{(0)}_{gg\to {t\bar t}} 
  \Biggr\{ 
         96 \* \ln^2 \beta
       - 9.5165 \* \ln \beta
       + 35.322 + 5.1698 \* {1 \over \beta}
    \Biggr\}\, ,
\\
\label{eq:sgg20-num}
  \hat{\sigma}^{(2)}_{gg\to {t\bar t}} &=& 
  \hat{\sigma}^{(0)}_{gg\to {t\bar t}} 
  \Biggr\{ 
         4608 \* \ln^4 \beta
       - 1894.9 \* \ln^3 \beta
       + \left(  - 3.4811 + 496.30 \* {1 \over \beta} \right) \* \ln^2 \beta
\nonumber \\ &&\qquad\qquad
       + \left( 3144.4 + 321.17 \* {1 \over \beta} \right) \* \ln \beta
       + 68.547 \* {1 \over \beta^2} - 196.93 \* {1 \over \beta} + C^{(2)}_{gg}
    \Biggr\}\, ,
\end{eqnarray}
The terms proportional to inverse powers of $\beta$ correspond to the Coulomb
corrections and the presently unknown two-loop constants $C^{(2)}_{q{\bar q}}$ and $C^{(2)}_{gg}$ are set to zero.
For reference, we have also repeated the well-known NLO results Eq.~(\ref{eq:sqq10-num}) and Eq.~(\ref{eq:sgg10-num}). 
The lengthy analytical results (containing the explicit dependence on the color factors $C_A, C_F$ and on $n_f$) 
are given in the Appendix, Eqs.~(\ref{eq:sqq10})--(\ref{eq:sgg20}).
We define a \NNLO\ cross section to be used in this paper as the sum of exact NLO result and the two-loop
contribution of Eqs.~(\ref{eq:sqq20-num}) and (\ref{eq:sgg20-num}) for all {\it scale independent} terms.
All {\it scale dependent} terms at NNLO accuracy are long known
exactly~\cite{Kidonakis:2001nj} 
as they can be easily constructed from the lower orders convoluted with the appropriate splitting functions.
We use the exact result of Ref.~\cite{Kidonakis:2001nj} for the $\mu$-dependence at two loops. 
As emphasized several times, the $qg$- and ${\bar q}g$-contributions are small at Tevatron and LHC and 
we simply keep them at NLO here.

\begin{figure}[htb]
  \begin{center}
    \includegraphics[width=0.49\textwidth]{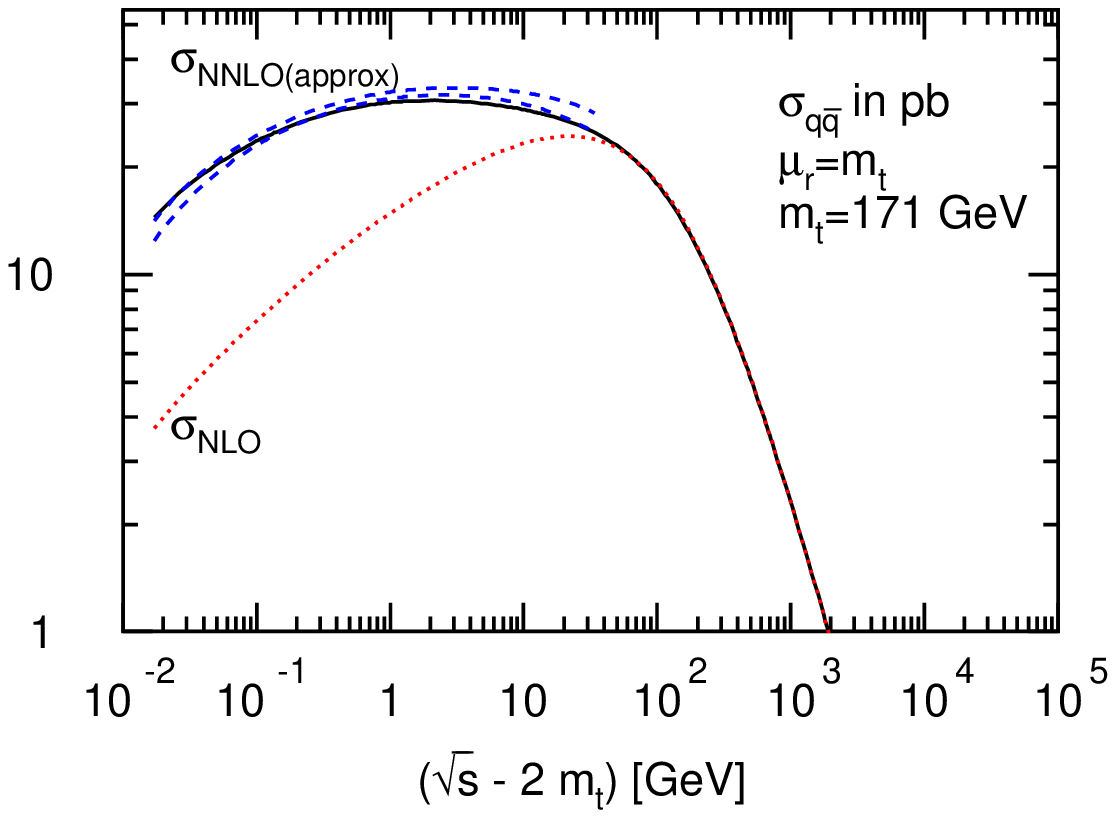}
    \includegraphics[width=0.49\textwidth]{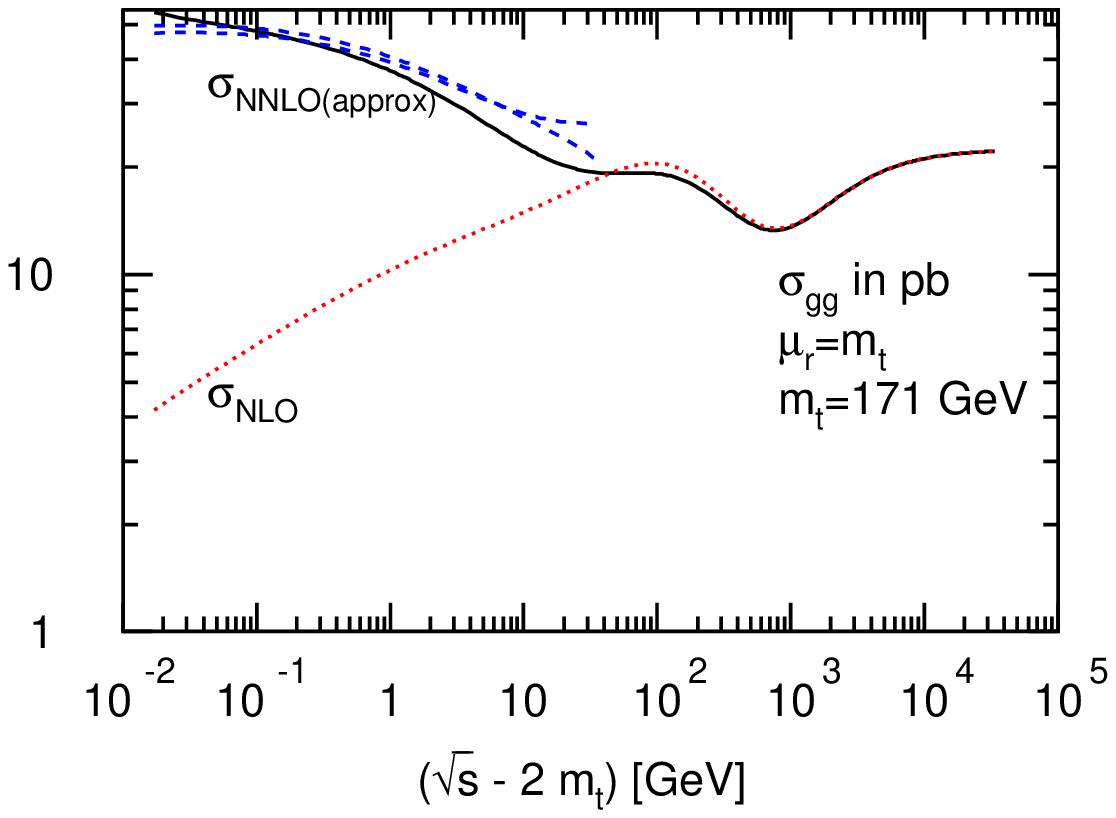}
    \vspace*{-1mm}
    \caption{ \small
      \label{pic:nnlo-partonic-crs}
      The partonic cross sections for the processes 
      ${q\bar q} \to {t\bar t}$ (left) and $gg \to {t\bar t}$ (right)
      in pb, and for
      $\mu = \mt=171$~GeV. 
      The dotted lines are the exact NLO 
      result~\cite{Nason:1988xz,Beenakker:1989bq} and the solid lines 
      correspond to the \NNLO\ result of this paper.
      The dashed lines close by denote the previous 
      approximations of~\cite{Kidonakis:2001nj}.
      }
    \vspace*{2mm}
  \end{center}
\end{figure}
In Fig.~\ref{pic:nnlo-partonic-crs} we display the \NNLO\ results for the partonic cross sections 
as calculated in Eqs.~(\ref{eq:sqq20-num}) and (\ref{eq:sgg20-num}). 
We plot $\hat{\sigma}_{{q\bar q} \to {t\bar t}}$ (left) and $\hat{\sigma}_{gg \to {t\bar t}}$ (right) 
again as a function of the distance from threshold at $\sqrt{\shat} = 2 \mt$ 
(solid lines in Fig.~\ref{pic:nnlo-partonic-crs}). 
This is useful to asses the question what improvements can be expected from a full NNLO calculation. 
Given that the $K$-factor of the NLO result is of moderate size at large partonic energies 
one can expect that the full NNLO corrections should give only small corrections in this region of phase space. 
On the other hand the $K$-factor of the NLO correction becomes large in the threshold region 
(i.e. $\sqrt{\shat} - 2 \mt\lsim {\cal O}(50)$~GeV), where finally perturbation theory breaks down. 
In this region the corrections are dominated by the large logarithms in
$\beta$ together with the Coulomb corrections which are correctly described 
by Eqs.~(\ref{eq:sqq20-num}) and (\ref{eq:sgg20-num}). 
In other words, where we expect large corrections from the full NNLO
QCD calculation our $\sigmaNNLO$ should provide a good estimate. This
is further supported by Ref.~\cite{Dittmaier:2007wz} where the
next-to-leading order corrections to top-quark pair production
together with an additional jet have been calculated. This
contribution represents part of the NNLO corrections for inclusive
top-quark pair production. In Ref.~\cite{Dittmaier:2007wz} it was
found that for $\mu = \muf=\mur=\mt$ the NLO corrections to 
$t\bar t + \mbox{1-jet}$ production are almost zero. This is a further indication
that the hard corrections to the inclusive top-quark pair production
at NNLO are indeed small.
Moreover, as mentioned above, we have further improved $\sigmaNNLO$ by incorporating 
the complete scale dependence at NNLO which is already known exactly~\cite{Kidonakis:2001nj}.
To that end let us quantify in detail once more the range of validity for the soft gluon approximations. 
We show in Tab.~\ref{tab:betatable} the numerical size of the individual logarithms and powers in $\beta$ 
in Eqs.~(\ref{eq:sqq10-num})--(\ref{eq:sgg20-num}) when evaluated at a given parton energy $\shat$. 
As a consequence of the moderate size of the expansion coefficients in Eqs.~(\ref{eq:sqq10-num})--(\ref{eq:sgg20-num}), 
we clearly see the good convergence properties of the logarithmic expansion up energies 
$\sqrt{\shat} - 2 \mt\lsim {\cal O}(50)$~GeV.
Beyond that value the soft logarithms (being proportional to the Born cross section) smoothly vanish.
\begin{table}[htbp]
  \begin{center}
    \leavevmode
     \begin{tabular}[h]{r|r|r|r|r|r|r|r}
 ${\sqrt{\shat}} - 2\mt$~[GeV] & $\ln^4\beta$~~~ & 
 $\ln^3\beta$~~~& $\ln^2\beta$~~~& $\ln\beta$~~~&
 $\beta^{-2}$~~~& $\beta^{-1}$~~~& $\beta$~~~ \\
 \hline
        0.1  ~~~~~~~~&   191.9844  &   -51.5762  &    13.8558  &    -3.7223  &  1710.7500  &    41.3612  &     0.0242  \\
        0.5  ~~~~~~~~&    72.5503  &   -24.8588  &     8.5176  &    -2.9185  &   342.7502  &    18.5135  &     0.0540  \\
        1.0  ~~~~~~~~&    43.8302  &   -17.0345  &     6.6204  &    -2.5730  &   171.7504  &    13.1054  &     0.0763  \\
        5.0  ~~~~~~~~&     9.9709  &    -5.6111  &     3.1577  &    -1.7770  &    34.9518  &     5.9120  &     0.1691  \\
       10.0  ~~~~~~~~&     4.3130  &    -2.9928  &     2.0768  &    -1.4411  &    17.8536  &     4.2254  &     0.2367  \\
       50.0  ~~~~~~~~&     0.2628  &    -0.3671  &     0.5127  &    -0.7160  &     4.1870  &     2.0462  &     0.4887  \\
      100.0  ~~~~~~~~&     0.0434  &    -0.0951  &     0.2084  &    -0.4565  &     2.4919  &     1.5786  &     0.6335  \\
      500.0  ~~~~~~~~&     0.0001  &    -0.0007  &     0.0081  &    -0.0901  &     1.1976  &     1.0943  &     0.9138  \\
 \hline
 \end{tabular}

  \caption{ \small
      \label{tab:betatable}
      Numerical values of the individual powers of $\ln \beta$ and $\beta$
      for various distances from threshold $\sqrt{\shat} - 2 \mt$ for $\mt=171$~GeV 
      as entering in Eqs.~(\ref{eq:sqq10-num})--(\ref{eq:sgg20-num}).
    }
  \end{center}
\end{table}

In addition we show in Fig.~\ref{pic:nnlo-partonic-crs} also previous approximations 
to the NNLO correction from Ref.~\cite{Kidonakis:2001nj} (dashed lines) 
employing two distinct differential kinematics to define the partonic threshold. 
They agree with our \NNLO\ corrections for $\sqrt{\shat} - 2 \mt\lsim 30$~GeV 
if Eqs.~(\ref{eq:sqq20-num}) and (\ref{eq:sgg20-num}) are truncated to the first three powers in $\ln \beta$.
However, at higher partonic center-of-mass energies the results of
Ref.~\cite{Kidonakis:2001nj} receive large 
numerical contributions from sub-leading terms and become unreliable 
(see in particular Fig.~\ref{pic:nnlo-partonic-crs} on the right).

We would also like to point out that there is a discrepancy between the
NLL resummed cross section of Ref.~\cite{Bonciani:1998vc} and Eq.~(\ref{eq:defsigmares}) 
and the fixed order NNLO approximation discussed here. 
In particular for the gluon fusion channel in Fig.~\ref{pic:r-partonic-crs} (right) and Fig.~\ref{pic:nnlo-partonic-crs} (right) 
the numerical differences between $\sigmaRES$ and $\sigmaNNLO$ are rather large.
The all-order NLL resummed cross section $\sigmaRES$ is significantly smaller 
than its expansion to second-order Eq.~(\ref{eq:sgg20-num}) or the
corresponding result of Ref.~\cite{Kidonakis:2001nj}, the latter two both being consistent with each other.
We can attribute this difference to the following fact: For $t{\bar t}$-hadro-production 
the Born cross section exhibits simple (although non-trivial) $N$-dependence 
and the resummed cross section $\sigmaRES$ in Mellin space (as
implemented in Ref.~\cite{Bonciani:1998vc} and Eq.~(\ref{eq:defsigmares})) 
contains products of $N$-dependent functions. 
This is unlike other cases considered in the literature, where the
Born terms have always been proportional to a delta-function, 
i.e. $\delta(1-x)$ for DIS, Drell-Yan or Higgs production.
In momentum space products of $N$-dependent functions correspond to convolutions, 
which induce formally sub-leading but numerically large corrections in the resummed result.
Eventually, this leads to the observed suppression in Fig.~\ref{pic:r-partonic-crs}.
Most likely this large discrepancy will also persist when comparing $\sigmaRES$ to a full NNLO QCD calculation, 
or upon matching the latter to a resummed cross section at NNLL accuracy along the lines of Eq.~(\ref{eq:defsigmares}).
This fact has to be kept in mind when using $\sigmaRES$ for predictions at LHC, where the $gg$-channel dominates.
As mentioned above, for any finite order expansion in $\alpha_s$, 
there is no ambiguity in performing the inverse Mellin transformation
analytically up to power suppressed terms in $N$ or, equivalently in $(1-\rho)$.

\medskip

We are now in a position to present the new results for the top-quark cross section $\sigma_{pp \to t{\bar t}X}$ 
at \NNLO\ as defined below Eq.~(\ref{eq:sgg20-num}) including the exact scale dependence.
We also quote the corresponding uncertainty according to Eq.~(\ref{eq:range}).
In our study we use the same PDFs as in 
Tabs.~\ref{tab:CTEQ6.5-MATCHED-Tevatron}--\ref{MRST06nnlo-MATCHED-LHC} above.
\begin{figure}[htb]
  \begin{center}
    \includegraphics[width=0.65\textwidth]{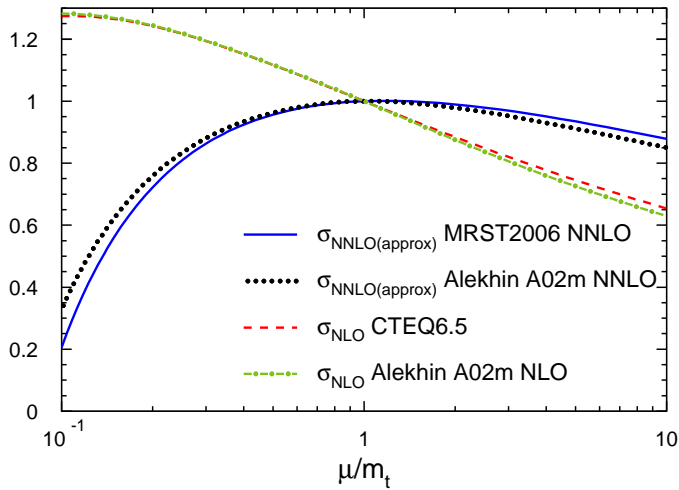}
    \vspace*{-1mm}
    \caption{ \small
      \label{pic:scaledep-lhc}
      Comparison of the scale dependence of \sigmaNNLO\ with 
      PDF set MRST-2006~NNLO~\cite{Martin:2007bv} and
      \sigmaNLO\ with PDF set CTEQ6.5~\cite{Tung:2006tb}. For
      comparison we show also the corresponding results for
      the Alekhin set of PDFs~\cite{Alekhin:2005gq}.
      The cross sections are normalized to the value at $\mu=\mt$.
      }
    \vspace*{2mm}
  \end{center}
\end{figure}
In Fig.\ref{pic:scaledep-lhc} the scale dependence for \sigmaNLO\  
and \sigmaNNLO\ is shown. For the \sigmaNNLO\  we use the PDF set
MRST-2006 NNLO  while for \sigmaNLO\  the PDF set CTEQ6.5 is used. To
become less sensitive to different normalizations we normalise the
curves to the central value $\sigma(\mu=\mt)$. In addition we show
also the results obtained by using the Alekhin PDF set
\cite{Alekhin:2005gq} available in NLO and NNLO accuracy. 
After normalisation one can see that the two curves for the NLO
predictions agree rather well as one might expect. For the two NNLO
curves the agreement is less good in particular for extreme values of
the scale. The origin of the minor discrepancy might be attributed to
slightly different input densities at low scale. 
Compared to the NLO results the scale dependence of
\sigmaNNLO\ is improved. In particular we find a plateau at $\mu=\mt$.
Note that due to the different shape of the NNLO curve the uncertainty estimate of
Eq.~(\ref{eq:range}) has to be adapted here. 
Restricting the scale to the interval $[\mt/2, 2\mt]$ the residual scale dependence of
$\sigmaNNLO$ is reduced to a few percent.

\begin{figure}[htb]
  \begin{center}
  \includegraphics[width=0.49\textwidth]{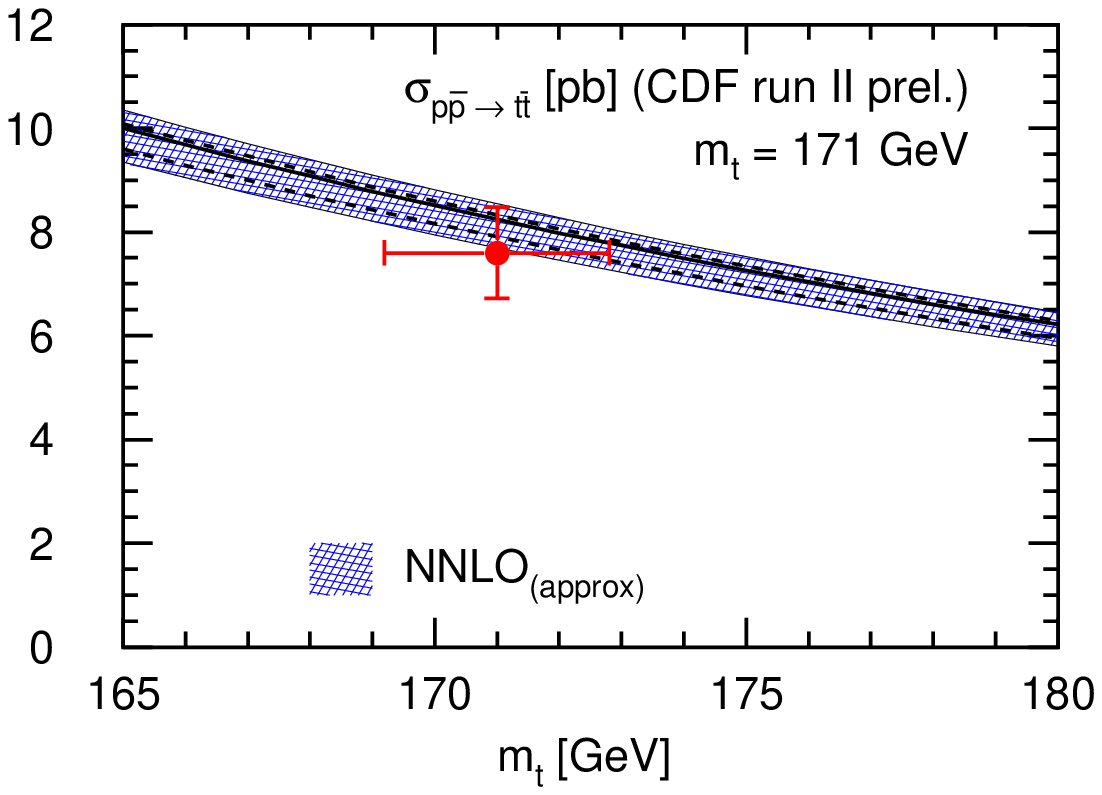}\hfill
  \includegraphics[width=0.49\textwidth]{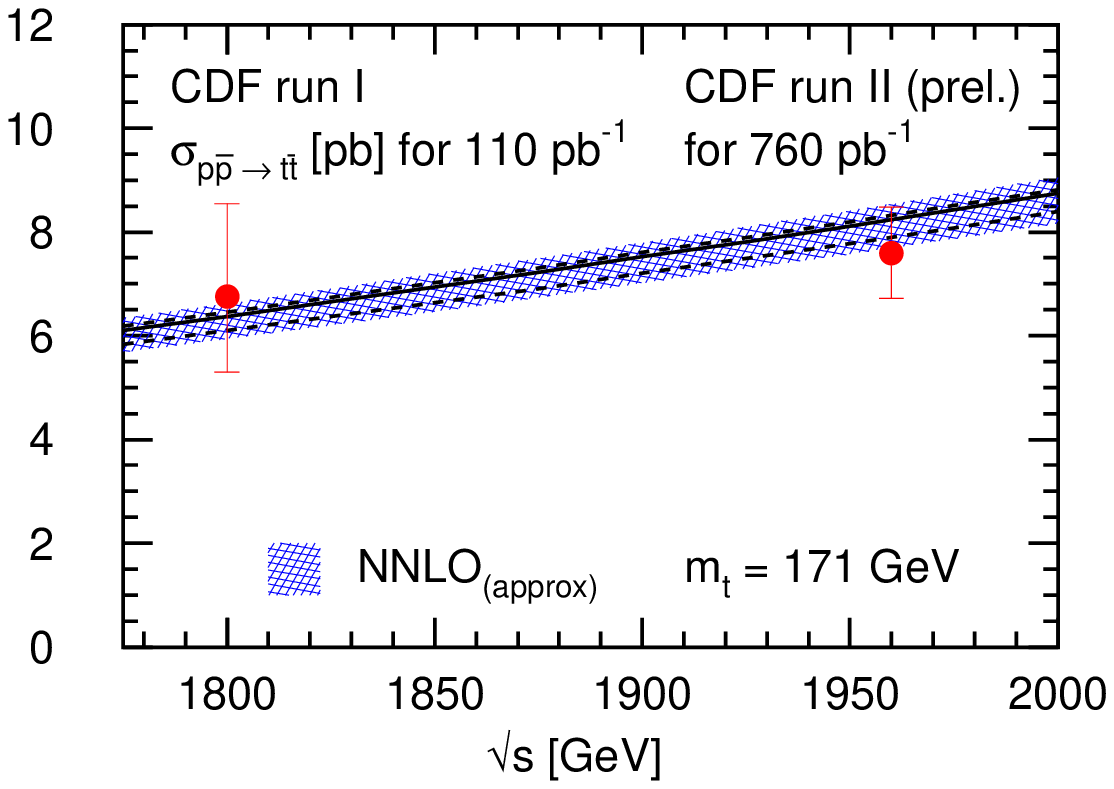}
  \vspace*{-1mm}
  \caption{ \small
    \label{pic:ttbar-total-tev}
    The \NNLO\ QCD prediction for the $t{\bar t}$ total cross section at
    Tevatron and CDF data~\cite{cdf:2006note} with $\mt=171$~GeV -- 
    as functions of $\mt$ for $\sqrt{ \shad}=1.96$~TeV (left) and 
    of $\sqrt {\shad}$ (right).
    The solid line is the central value for $\mu=\mt$, the dashed 
    lower and upper
    lines correspond to $\mu=2\mt$ and $\mu=\mt/2$, respectively.
    The band denotes the total uncertainty that is the
    uncertainty due to scale variations and the PDF uncertainty of
    the MRST-2006~NNLO set~\cite{Martin:2007bv} combined together according to
    Eq.~(\ref{eq:range}).
    }
\vspace*{2mm}
  \end{center}
\end{figure}

\begin{figure}[htb]
  \begin{center}
  \includegraphics[width=0.49\textwidth]{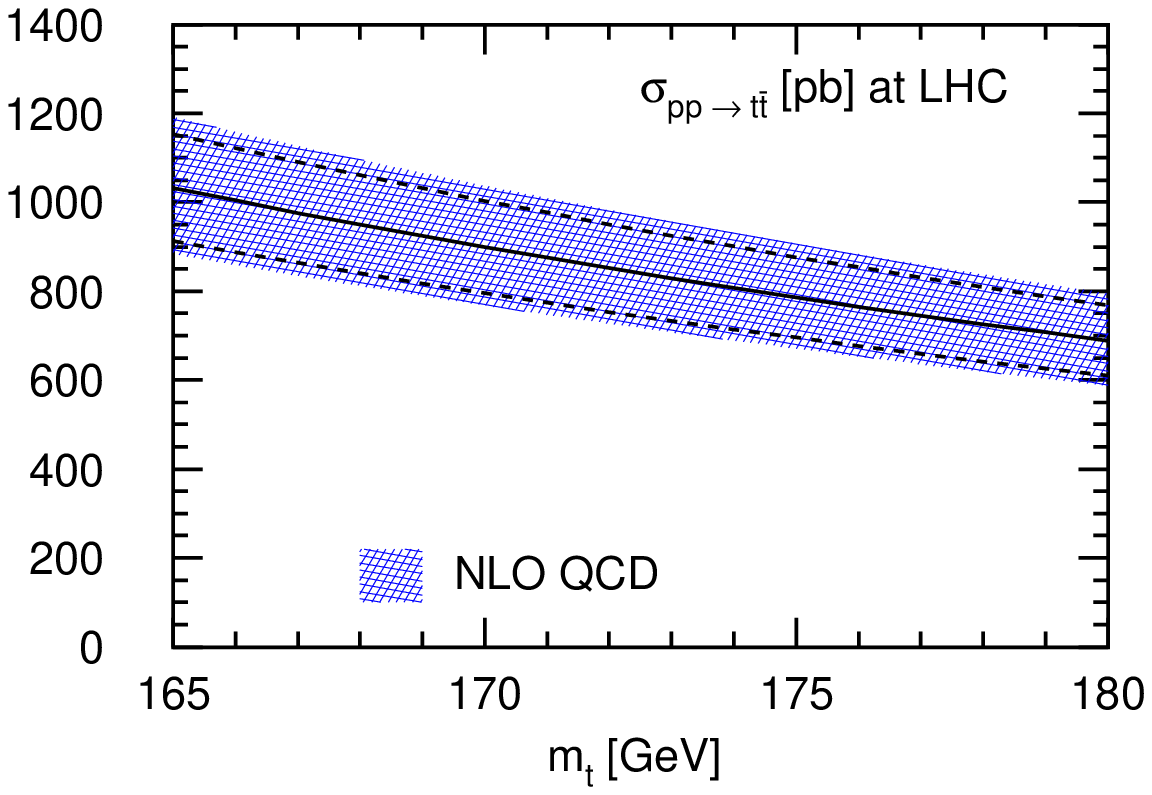}
  \includegraphics[width=0.49\textwidth]{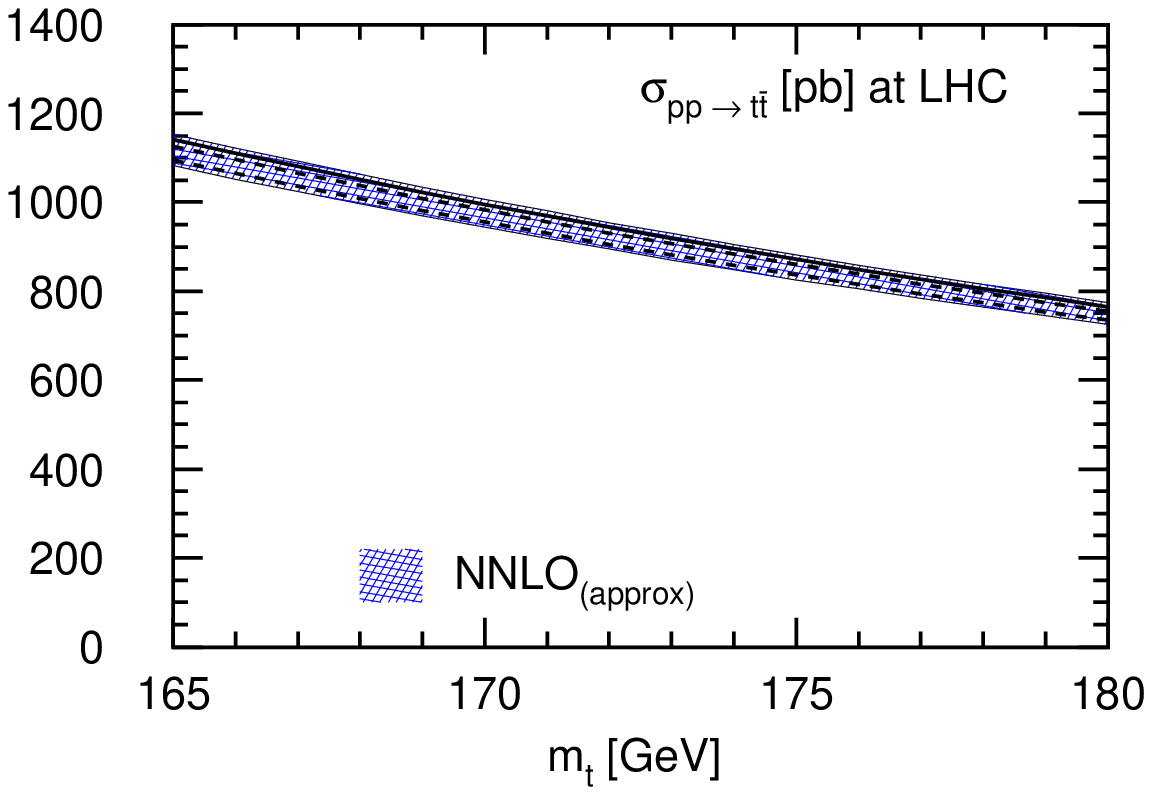}
  \vspace*{-1mm}
  \caption{ \small
    \label{pic:ttbar-total-lhc}
    The \NNLO\ QCD prediction for the $t{\bar t}$ total cross section at
    LHC as functions of $\mt$ for $\sqrt{ \shad}=14$~TeV (right).
    The solid line is the central value for $\mu=\mt$, the dashed
    lower and upper
    lines correspond to $\mu=2\mt$ and $\mu=\mt/2$, respectively.
    The band denotes the total uncertainty that is the
    uncertainty due to scale variations and the PDF uncertainty of
    the  MRST-2006 NNLO set~\cite{Martin:2007bv}.
    For comparison the left plot shows 
    the corresponding prediction at NLO accuracy using the 
    PDF set CTEQ6.5 \cite{Tung:2006tb}.
  }
\vspace*{2mm}
  \end{center}
\end{figure}

In Fig.~\ref{pic:ttbar-total-tev} we plot the total cross section at Tevatron
and display also CDF data~\cite{cdf:2006note} with $\mt=171$~GeV.
The corresponding cross section values are given in 
Tabs.~\ref{tab:CTEQ6.5-NNLO-Tevatron} and \ref{tab:MRST06nnlo-NNLO-Tevatron}.
By using \sigmaNNLO\ the residual scale dependence is reduced to
3\%. Compared to \sigmaNLO\ and \sigmaRES\ presented in the previous
section this corresponds to a reduction by a factor of 2. The data
points nicely agree with the theoretical prediction. The overall
uncertainty of the the theoretical prediction is about 8\% for CTEQ6.5
and 6\% for MRST-2006 NNLO --- an accuracy unlikely to be reached
at the Tevatron experiments. Note that the smaller PDF uncertainty
obtained when using the MRST-2006 NNLO PDF set is due to a different
convention used by the MRST collaboration to define the PDF
uncertainty.
In Fig. \ref{pic:ttbar-total-lhc} (right) we show \sigmaNNLO\ for the LHC. 
Again we observe a drastic reduction of the scale 
uncertainty compared to \sigmaNLO\ and \sigmaRES. For comparison we
show also \sigmaNLO\ in Fig. \ref{pic:ttbar-total-lhc} (left). 
The corresponding cross section values are listed in Tabs.~\ref{tab:CTEQ6.5-NNLO-LHC} and \ref{tab:MRST06nnlo-NNLO-LHC}.
Apart from reducing the scale dependence the \sigmaNNLO\ leads only to a small
shift of a few percent in the central value for the cross section prediction. The
\sigmaNNLO\ band is well contained in the \sigmaNLO\ band. 
So perturbation theory seems to be well behaved and under control. 
The overall uncertainty is about 6\% for the CTEQ6.5 PDF set and about 4\% for the 
MRST-2006 NNLO set.

The numbers quoted in Tabs.~\ref{tab:CTEQ6.5-NNLO-Tevatron}--\ref{tab:MRST06nnlo-NNLO-LHC} 
represent presently the best estimates for the top-quark production cross section at Tevatron and LHC 
(see the Appendix for additional information on the individual PDFs and their eigenvalues).
It should be kept in mind, though, that there is an intrinsic uncertainty in the
central value at $\mu = \mt$ of our \NNLO\ result due to neglected power corrections 
in $\beta \sim (1-\rho)$ away from threshold. 
However, due to the steeply falling parton flux (see Figs.~\ref{pic:LHCpdf}, \ref{pic:TEVpdf}), 
the numerical impact of these contributions is much suppressed. 


\begin{table}[htbp]
  \begin{center}
    \leavevmode
    \begin{tabular}{c|llc|llc|llc} \hline
 &\multicolumn{3}{|c|}{only scale uncertainty}&\multicolumn{3}{|c|}{only pdf uncertainty}&\multicolumn{3}{|c}{total uncertainty}\\ 
m &min&max&$\delta [\%]$&min&max&$\delta [\%]$&min&max&$\delta [\%]$\\ \hline\hline
165&9.26&9.66&3&9.04&10.1&6&8.73&10.1&8\\
166&8.97&9.36&3&8.76&9.82&6&8.46&9.87&8\\
167&8.68&9.07&3&8.49&9.51&6&8.2&9.57&8\\
168&8.41&8.79&3&8.22&9.22&6&7.94&9.27&8\\
169&8.15&8.52&3&7.97&8.93&6&7.7&8.98&8\\
170&7.9&8.26&3&7.73&8.65&6&7.46&8.7&8\\
171&7.65&8.01&3&7.49&8.38&6&7.23&8.44&8\\
172&7.42&7.76&3&7.26&8.12&6&7.01&8.18&8\\
173&7.19&7.53&3&7.04&7.87&6&6.8&7.93&8\\
174&6.97&7.3&3&6.83&7.63&6&6.59&7.69&8\\
175&6.76&7.08&3&6.62&7.4&6&6.39&7.45&8\\
176&6.55&6.87&3&6.43&7.17&6&6.2&7.23&8\\
177&6.36&6.66&3&6.23&6.96&6&6.01&7.01&8\\
178&6.16&6.46&3&6.05&6.75&6&5.83&6.8&8\\
179&5.98&6.27&3&5.87&6.54&6&5.66&6.6&8\\
180&5.8&6.08&3&5.69&6.35&6&5.49&6.4&8\\
\hline
\end{tabular}

  \end{center}
  \caption{ \small
      \label{tab:CTEQ6.5-NNLO-Tevatron}
      The cross section $\sigmaNNLO$ as derived in this paper 
      in pb for various values of the top-quark mass $\mt$ at Tevatron
      ($\sqrt{ \shad}=1.96$~TeV) using the CTEQ6.5 PDF set~\cite{Tung:2006tb}.
    }
\end{table}
\begin{table}[htbp]
  \begin{center}
    \leavevmode
    \begin{tabular}{c|llc|llc|llc} \hline
 &\multicolumn{3}{|c|}{only scale uncertainty}&\multicolumn{3}{|c|}{only pdf uncertainty}&\multicolumn{3}{|c}{total uncertainty}\\ 
m &min&max&$\delta [\%]$&min&max&$\delta [\%]$&min&max&$\delta [\%]$\\ \hline\hline
165&9.59&10&3&9.73&10.2&3&9.34&10.3&6\\
166&9.28&9.76&3&9.42&9.94&3&9.04&10&6\\
167&8.99&9.45&3&9.12&9.62&3&8.75&9.7&6\\
168&8.7&9.16&3&8.83&9.31&3&8.47&9.39&6\\
169&8.42&8.87&3&8.55&9.02&3&8.2&9.1&6\\
170&8.16&8.59&3&8.28&8.73&3&7.94&8.81&6\\
171&7.9&8.32&3&8.02&8.46&3&7.69&8.53&6\\
172&7.65&8.06&3&7.77&8.19&3&7.45&8.27&6\\
173&7.41&7.81&3&7.53&7.93&3&7.22&8.01&6\\
174&7.18&7.57&3&7.29&7.69&3&6.99&7.76&6\\
175&6.95&7.34&3&7.07&7.45&3&6.77&7.52&6\\
176&6.74&7.11&3&6.85&7.22&3&6.57&7.29&6\\
177&6.53&6.89&3&6.64&6.99&3&6.36&7.07&6\\
178&6.33&6.68&3&6.43&6.78&3&6.17&6.85&6\\
179&6.13&6.48&3&6.24&6.57&3&5.98&6.64&6\\
180&5.95&6.28&3&6.05&6.37&3&5.8&6.44&6\\
\hline
\end{tabular}

  \end{center}
  \caption{ \small
      \label{tab:MRST06nnlo-NNLO-Tevatron}
      Same as in Tab.~\ref{tab:CTEQ6.5-NNLO-Tevatron} using the MRST-2006 NNLO PDF set~\cite{Martin:2007bv}.
    }
\end{table}
\begin{table}[htbp]
  \begin{center}
    \leavevmode
    \begin{tabular}{c|llc|llc|llc} \hline
 &\multicolumn{3}{|c|}{only scale uncertainty}&\multicolumn{3}{|c|}{only pdf uncertainty}&\multicolumn{3}{|c}{total uncertainty}\\ 
m &min&max&$\delta [\%]$&min&max&$\delta [\%]$&min&max&$\delta [\%]$\\ \hline\hline
165&1035&1082&3&1048&1117&4&1003&1117&6\\
166&1007&1052&3&1019&1086&4&975&1086&6\\
167&979&1024&3&991&1056&4&948&1056&6\\
168&953&996&3&964&1028&4&922&1028&6\\
169&927&969&3&937&1000&4&897&1000&6\\
170&902&943&3&912&973&4&873&973&6\\
171&878&917&3&887&947&4&849&947&6\\
172&855&893&3&863&922&4&827&922&6\\
173&832&869&3&840&898&4&805&898&6\\
174&810&846&3&818&874&4&783&874&6\\
175&789&824&3&796&851&4&762&851&6\\
176&768&802&3&775&829&4&742&829&6\\
177&748&781&3&755&808&4&723&808&6\\
178&729&761&3&735&787&4&704&787&6\\
179&710&741&3&716&767&4&686&767&6\\
180&692&722&3&698&747&4&668&747&6\\
\hline
\end{tabular}

  \end{center}
  \caption{ \small
      \label{tab:CTEQ6.5-NNLO-LHC}
      The cross section $\sigmaNNLO$ as derived in this paper 
      in pb for various values of the top-quark mass $\mt$ at LHC 
      ($\sqrt{ \shad}=14$~TeV) using the CTEQ6.5 PDF set~\cite{Tung:2006tb}.
    }
\end{table}
\begin{table}[htbp]
  \begin{center}
    \leavevmode
    \begin{tabular}{c|llc|llc|llc} \hline
 &\multicolumn{3}{|c|}{only scale uncertainty}&\multicolumn{3}{|c|}{only pdf uncertainty}&\multicolumn{3}{|c}{total uncertainty}\\ 
m &min&max&$\delta [\%]$&min&max&$\delta [\%]$&min&max&$\delta [\%]$\\ \hline\hline
165&1094&1141&3&1128&1154&2&1082&1154&4\\
166&1064&1110&3&1097&1122&2&1052&1122&4\\
167&1035&1080&3&1067&1092&2&1024&1092&4\\
168&1008&1050&3&1038&1063&2&996&1063&4\\
169&981&1022&3&1010&1034&2&969&1034&4\\
170&955&995&3&983&1007&2&943&1007&4\\
171&929&969&3&957&980&2&918&980&4\\
172&905&943&3&932&954&2&894&954&4\\
173&881&918&3&907&929&2&871&929&4\\
174&858&894&3&883&905&2&848&905&4\\
175&836&871&3&860&882&2&826&882&4\\
176&814&848&3&838&859&2&804&859&4\\
177&793&826&3&816&837&2&783&837&4\\
178&773&805&3&795&815&2&763&815&4\\
179&753&785&3&775&795&2&744&795&4\\
180&734&765&3&755&774&2&725&774&4\\
\hline
\end{tabular}

  \end{center}
  \caption{ \small
      \label{tab:MRST06nnlo-NNLO-LHC}
      Same as in Tab.~\ref{tab:CTEQ6.5-NNLO-LHC} using the MRST-2006 NNLO PDF set~\cite{Martin:2007bv}.
    }
\end{table}

\newpage
\section{Conclusions}
\label{sec:conclusions}

In this article we have summarized the present knowledge on theory predictions 
for the top-quark pair production cross section at Tevatron and LHC. 
We have taken some care to quantify the sensitivity of the total cross section
to soft gluon emission and large Sudakov type logarithms. 
As is well known, top-quark pair production at Tevatron is largely dominated by 
parton kinematics close to threshold, thus approximations based on soft gluon
resummation should provide an excellent description.
At LHC we find that soft gluon emission near threshold is less dominant, 
but contributes still a numerically sizable fraction to the total cross section. 
Thus, soft gluon effects in $t{\bar t}$-production are still rather prominent at LHC as well.

We have updated the NLL resummed cross section as defined in
Ref.~\cite{Bonciani:1998vc,Cacciari:2003fi} 
using recent PDFs.
Furthermore, we have extended the resummed predictions to NNLL accuracy and we have derived approximate NNLO cross sections
which are exact to all powers in $\ln \beta$ at two loops.
Together with the exact NNLO scale dependence (and including the two-loop Coulomb corrections) 
our result for $\sigmaNNLO$ represents the best present estimate for
hadro-production of top-quark pairs, both at Tevatron and LHC. As mentioned earlier we believe
that hard corrections at the NNLO level are small. This is supported by
the explicit findings of Ref.~\cite{Dittmaier:2007wz}.
We have found for the NNLL resummed cross section and the finite order
expansion good apparent convergence properties. 
Moreover, the stability of the total cross section with respect to scale variations is 
much improved by our \NNLO\ result. 

In closing let us briefly comment on ideas to use top-quark pair production as an additional calibration 
process for the parton luminosity at LHC~\cite{Nadolsky:2008zw}.
This could become feasible because the PDF dependence of $t{\bar t}$-production at LHC
is anti-correlated with $W/Z$-boson production ({\it the} standard candle
process at LHC, see e.g.~\cite{Dittmar:1997md,Dittmar:2005ed}) 
and correlated with Higgs boson production, especially for larger Higgs masses.
It has been noted however, that the NLO theory predictions to the top-quark cross 
section are not accurate enough.
We are confident that the \NNLO\ results of this present paper provide a step in the right 
direction by further constraining the theory uncertainties for this important process.

The complete NNLO QCD predictions for heavy-quark hadro-production do not only require 
the hard scattering cross section but also the evolution of the parton densities 
to be performed at the same order employing the NNLO splitting functions~\cite{Moch:2004pa,Vogt:2004mw}.
In addition we remark, that the present accuracy on the gluon PDF in the
medium $x$ range of interest for top-quark production at LHC is well constrained by 
from DIS data for structure functions in $e^\pm p$-scattering from
HERA and evolution, leading to the rather small uncertainty of 3\% (see Fig.~\ref{pic:LHCpdf}) at small energies.
Thus, for top-quark pair production to become a standard candle process at LHC
similar to $W/Z$ gauge boson production and to become competitive with DIS data 
also an experimental accuracy much better than the currently
quoted~\cite{atlas:1999tdr2} value of 10\% will be needed.

\subsection*{Acknowledgements}
We would like to thank S.~Alekhin, W.~Bernreuther, and E.~Laenen for 
stimulating discussions. 
S.M. is supported by the Helmholtz Gemeinschaft under contract VH-NG-105 
and P.U. is a Heisenberg fellow of Deutsche Forschungsgemeinschaft (DFG).
This work is also partly supported by DFG in SFB/TR 9 
and the hospitality of the Galileo-Galilei Institute in Florence 
is gratefully acknowledged.

{\it Note added.}--- Shortly after submitting the preprint version of this article, the related
work~\cite{Cacciari:2008zb} appeared, where a similar study has been performed.
In the meantime a tuned comparison of the resummed (NLL) partonic cross section
has been done and we have found good agreement, once it was taken into
account that the matching for the scheme $A=2$ implemented in~\cite{Cacciari:2008zb}
is slightly different from the original Ref.~\cite{Bonciani:1998vc} employed
in the present article. This minor scheme dependence accounts for the observed 
differences in the predictions for the total cross section
between~\cite{Cacciari:2008zb} and this article.

We would also like to thank J.~K\"uhn for pointing out an error in the Coulumb
contributions at two loops in Eqs.~(\ref{eq:sqq20-num}), (\ref{eq:sqq20-num}), 
and (\ref{eq:sqq20}), (\ref{eq:sgg20}) respectively, in the first version of this article.

\appendix

\renewcommand{\theequation}{\ref{sec:appA}.\arabic{equation}}
\setcounter{equation}{0}
\section{Useful formulae}
\label{sec:appA}

The Born cross section in the color-basis defined by color-singlet and
color-octet final states reads in momentum space with $\rho = 4\mt^2/s$ 
and $\beta = \sqrt{1 - \rho}$
\begin{eqnarray}
  \label{eq:Borncrs}
  \hat{\sigma}^{(0)}_{q{\bar q},\, 1}(\rho) &=& 0\, ,
\nonumber \\
  \hat{\sigma}^{(0)}_{q{\bar q},\, 8}(\rho) &=& {\alpha_s^2 \over \mt^2} 
  {\pi \over 12}\, {\cf \over \nc}\, \beta \rho \, ( 2 + \rho)
\, ,
\nonumber \\
  \hat{\sigma}^{(0)}_{gg,\, 1}(\rho) &=& {\alpha_s^2 \over \mt^2} 
  {\pi \over 16}\, {1 \over \nc (\nc^2-1)}\, \beta \rho \, 
  \left\{ ( 4 + 4 \rho - 2 \rho^2)\, {1 \over \beta}\, \ln {1+\beta \over 1-\beta} - 4 - 4 \rho \right\} 
\, ,
\nonumber \\
  \hat{\sigma}^{(0)}_{gg,\, 8}(\rho) &=& {\alpha_s^2 \over \mt^2} 
  {\pi \over 24}\, {1 \over \nc^2-1}\, \beta \rho \, \Biggl\{
  12 \cf \left[ ( 2 + 2 \rho - \rho^2)\, {1 \over \beta}\, \ln {1+\beta \over 1-\beta} - 2 - 2 \rho \right]
\nonumber \\
&&
  - \ca \left[ 6 ( 1 + \rho - \rho^2)\, {1 \over \beta}\, \ln {1+\beta \over 1-\beta} - 2 - \rho \right]
\Biggr\}
\, ,
\end{eqnarray}
where $C_A$ and $C_F$ are the usual color factors, with $\ca = N = 3$ 
and  $C_F = {1\over 2 N}(N^2-1) = 4/3$ in QCD. 
The Mellin moments as defined in Eq.~(\ref{eq:mellindef}) can be
easily computed, see e.g. Ref.~\cite{Bonciani:1998vc}.

Next we present the perturbative expansions for the anomalous dimensions 
$A_q$, $A_g$, $D_q$ and $D_g$ entering Eq.~(\ref{eq:GNresDYH}). 
We have for the quark case~\cite{Kodaira:1982nh,Moch:2004pa}
\begin{eqnarray}
\label{eq:Aqexp}
  A^{(1)}_q & \! = \! & 4\, C_F 
\nonumber \\
  A^{(2)}_q & \! = \! & 8\, C_F \left[ \left( \frac{67}{18}
     - \zeta_2 \right) C_A - \frac{5}{9}\,n_f \right] 
\nonumber \\
  A^{(3)}_q & \! = \! &
     16\, C_F \left[ C_A^{\,2} \,\left( \frac{245}{24} - \frac{67}{9}
     \: \zeta_2 + \frac{11}{6}\:\zeta_3 + \frac{11}{5}\:\zeta_2^2 \right)
   \: + \: C_F n_f\, \left( -  \frac{55}{24}  + 2\:\zeta_3
   \right) \right. 
\nonumber\\ & & \left. \mbox{} \qquad
   + \: C_A n_f\, \left( - \frac{209}{108}
         + \frac{10}{9}\:\zeta_2 - \frac{7}{3}\:\zeta_3 \right)
   \: + \: n_f^2 \left( - \frac{1}{27}\,\right) \right] \:\: ,
\end{eqnarray}
where $n_f$ denotes the number of effectively massless quark flavors 
and $\z2,\z3,\dots$ are the values of the Riemman zeta-function.
Likewise, the $D_q$ read 
\begin{eqnarray}
  D^{(1)}_q & = & 0\, ,
\nonumber \\
  D^{(2)}_q & = & C_F\, \left[
   C_A \left( - \frac{1616}{27} + \frac{176}{3}\,\zeta_2 + 56\,\zeta_3 \right)
   \: + \: n_f \left( \frac{224}{27} - \frac{32}{3}\,\zeta_2 \right)
   \right] \:\: .
\end{eqnarray}
All gluonic quantities are given by the simple relation
\begin{equation}
\label{eq:AgDg}
 A^{(i)}_g  =  {C_A \over C_F} \, A^{(i)}_q \, ,
 \qquad\qquad
 D^{(i)}_g  =  {C_A \over C_F} \, D^{(i)}_q \, .
\end{equation}

Here we summarize the functions $g^1_{ij}$, $g^2_{ij,\, I}$, $g^3_{ij,\, I}$
appearing in the resummed cross section Eq.(\ref{eq:GNexp})
to NNLL accuracy~\cite{Vogt:2000ci,Catani:2003zt,Moch:2005ba}.
Keeping the full dependence on $\mur$ and $\muf$, we have
\begin{eqnarray}
  \label{eq:g1res}
  g^1_{q{\bar q}} &=& 
          \Aq1 \*  (
            2
          - 2\*\ln(1-2\*\lambda)
          + \lambda^{-1} \* \ln(1-2\*\lambda) 
          )
\, ,
\\
  \label{eq:g2res}
  g^2_{q{\bar q},\, 1} &=& 
        \bigl(
            \Aq1 \* \beta_1 
          - \Aq2
        \bigr) \* (
            2\*\lambda
          + \ln(1-2\*\lambda)
          )
          + {1 \over 2} \* \Aq1 \* \beta_1 \* \ln^2(1-2\*\lambda)
       - {1 \over 2} \* \bigl( 4 \* \Aq1 \* \ec - \Dq1 \bigr) \* \ln(1-2\*\lambda)
\nonumber\\
&&\mbox{}
       + \Lqr \* \Aq1 \* \ln(1-2\*\lambda)
       + 2\* \Lfr \* \Aq1 \* \lambda
\, ,
\\
  g^2_{q{\bar q},\, 8} &=&
       g^2_{q{\bar q},\, 1} 
          - {1 \over 2} \* \ln(1 - 2 \* \lambda) \* \DQQ1
\, ,
\\
  \label{eq:g3res}
  g^3_{q{\bar q},\, 1} &=& 
         {1 \over 2} \* (
            \Aq1 \* \beta_2 
          - \Aq1 \* \beta_1^2 
          + \Aq2 \* \beta_1 
          - \Aq3 
         ) \*  \biggl(
            1
          + 2\*\lambda
          - {1 \over 1-2\*\lambda}
          \biggr)
\nonumber\\
&&\mbox{}
      +  \Aq1 \* \beta_1^2 \*  \biggl(
            {\ln(1-2\*\lambda) \over 1-2\*\lambda}
          + {1 \over 2} \* {\ln^2(1-2\*\lambda) \over 1-2\*\lambda}
          \biggr)
      + \biggl(
            \Aq1 \* \beta_2 
          - \Aq1 \* \beta_1^2 
         \biggr) \* \ln(1-2\*\lambda)
\nonumber\\
&&\mbox{}
       +  (
            2\*\Aq1 \* \beta_1 \* \ec 
          + \Aq2 \* \beta_1 
          - {1 \over 2}\*\Dq1 \* \beta_1 
          ) \* \biggl(
            1
          - {1 \over 1-2\*\lambda} 
          - {\ln(1-2\*\lambda) \over 1-2\*\lambda}
          \biggr)
\nonumber\\
&&\mbox{}
       - \biggl(
            \Aq1 \* \beta_2 
          + 2\*\Aq1  \* (\ecs + \z2) 
          + 2\*\Aq2 \* \ec
          - \Dq1 \* \ec
          - {1 \over 2}\*\Dq2 
         \biggr) \* \biggl(
            1
          - {1 \over 1-2\*\lambda}
          \biggr)
\nonumber\\
&&\mbox{}
       + \Lqr  \*  \biggl[
         (
            2\*\Aq1 \* \ec 
          - \Aq1 \* \beta_1 
          + \Aq2 
          - {1 \over 2}\*\Dq1 
          ) \*  \biggl(
            1 
          - {1 \over 1-2\*\lambda}
          \biggr)
       +  \Aq1 \* \beta_1 \* \biggl(
          {\ln(1-2\*\lambda) \over 1-2\*\lambda}
          \biggr)
          \biggl]
\nonumber\\
&&\mbox{}
       + 2 \* \Lfr \* \Aq2 \* \lambda 
       - {1 \over 2} \* \Lqrs \* \Aq1 \*  \biggl(
            1
          - {1 \over 1-2\*\lambda}
          \biggr)
       - \Lfrs \* \Aq1 \* \lambda 
\, ,
\\
  g^3_{q{\bar q},\, 8} &=&
       g^3_{q{\bar q},\, 1} 
       - {1 \over 2} \* (
            \DQQ2
          + 2 \* \DQQ1 \* \ec
          - \DQQ1 \* \b1 
          ) \* \biggl(
            1 
          - {1 \over (1 - 2 \* \lambda)} 
          \biggr)
          - {1 \over 2} \* {\ln(1 - 2 \* \lambda) \over (1 - 2 \* \lambda)} \* \DQQ1 \* \b1
\nonumber\\
&&\mbox{}
          + {1 \over 2} \* \Lqr \* \DQQ1 \* \biggl(
            1 
          - {1 \over (1 - 2 \* \lambda)} 
          \biggr)
\, ,
\end{eqnarray}
with $\gamma_e = 0.5772167$. 
The gluonic expressions $g^1_{gg}, g^2_{gg,\, I}$ and $g^3_{gg,\, I}$ are
obtained with the obvious replacements $A^{(i)}_q \to A^{(i)}_g$ and 
$D^{(i)}_q \to D^{(i)}_g$.
The dependence on $\beta_0$ is recovered by $A^{(i)} \to A^{(i)} / \beta_0^{\,i}$, 
$D^{(i)} \to D^{(i)} / \beta_0^{\,i}$, $\beta_i \to \beta_i/\beta_0^{\,i+1}$ 
and multiplication of $g^3_{ij,\, I}$ by $\beta_0$.
We also give explicit results for the matching functions $g^0_{ij,\, I}$ in Eq.(\ref{eq:sigmaNres}),
\begin{eqnarray}
  \label{eq:g0qq1}
  g^0_{q{\bar q},\, 1} &=& 
  0
\, ,
\\
  \label{eq:g0qq8}
  g^0_{q{\bar q},\, 8} &=&
  1 
       + a_s  \*  \Biggl\{
       \cf \* \biggl[
          - 64 
          + 8 \* \ecs
          + 4 \* \pi^2
          - 32 \* \lntwo
          + 8 \* \lntwos
       \biggr]
       + \ca \* \biggl[
          - 8
          + 4 \* \ec
          - 4 \* \lntwo
       \biggr]
          + 4 \* C^{(1)}_{q{\bar q}}
\nonumber\\
&&\mbox{}
       + \Lqr  \*  \cf \* \biggl[
           16
          - 8 \* \ec
          \biggr]
          \Biggr\}
\nonumber\\
&&\mbox{}
       + a_s^2 \* \Biggl\{
       \cf \* C^{(1)}_{q{\bar q}} \* \biggl[
          - 256
          + 32 \* \ecs
          + 16 \* \pi^2
          + 384 \* \lntwo
          - 64 \* \ec \* \lntwo
          - 256 \* \lntwos
          \biggr]
       + \cf \* \nf \* \biggl[
            {6976 \over 27}
\nonumber\\
&&\mbox{}
          - {448 \over 9} \* \z3
          - {224 \over 27} \* \ec
          - {80 \over 9} \* \ecs
          - {32 \over 9} \* \ect
          - {136 \over 9} \* \pi^2
          - {10240 \over 27} \* \lntwo
          + {160 \over 9} \* \ec \* \lntwo
          + {32 \over 3} \* \ecs \* \lntwo
\nonumber\\
&&\mbox{}
          + 16 \* \pi^2 \* \lntwo
          + {2368 \over 9} \* \lntwos
          - {32 \over 3} \* \ec \* \lntwos
          - {832 \over 9} \* \lntwoc
          \biggr]
       + \cfs \* \biggl[
          - 8192 
          + 3584  \* \z3
          - 512  \* \ecs
\nonumber\\
&&\mbox{}
          + 32  \* \ec^4
          + 256  \* \pi^2
          + 32  \* \pi^2 \* \ecs
          + 24  \* \pi^4
          + 12288  \* \lntwo
          - 5376  \* \z3 \* \lntwo
          + 1024  \* \ec \* \lntwo
\nonumber\\
&&\mbox{}
          - 128  \* \ect \* \lntwo
          - 768  \* \pi^2 \* \lntwo
          - 64  \* \pi^2 \* \ec \* \lntwo
          - 9728  \* \lntwos
          + 192  \* \ecs \* \lntwos
          + 608  \* \pi^2 \* \lntwos
\nonumber\\
&&\mbox{}
          + 6912  \* \lntwoc
          - 128  \* \ec \* \lntwoc
          - 2560  \* \lntwof
          \biggr]
       + \ca \* C^{(1)}_{q{\bar q}} \* \biggl[
          - 32
          + 16 \* \ec
          + 32 \* \lntwo
          \biggr]
\nonumber\\
&&\mbox{}
       + \ca \* \nf \* \biggl[
            {272 \over 9}
          - {40 \over 9} \* \ec
          - {8 \over 3} \* \ecs
          - {4 \over 3} \* \pi^2
          - {368 \over 9} \* \lntwo
          + {16 \over 3} \* \ec \* \lntwo
          + {64 \over 3} \* \lntwos
          \biggr]
\nonumber\\
&&\mbox{}
       + \ca \* \cf \* \biggl[
          - {55264 \over 27}
          + {7504 \over 9} \* \z3
          - {5296 \over 27} \* \ec
          - 56 \* \ec \* \z3
          - {40 \over 9} \* \ecs
          + {464 \over 9} \* \ect
          + {1276 \over 9} \* \pi^2
\nonumber\\
&&\mbox{}
          + 16 \* \pi^2 \* \ec
          - {8 \over 3} \* \pi^2 \* \ecs
          - {4 \over 3} \* \pi^4
          + {88192 \over 27} \* \lntwo
          - 112 \* \z3 \* \lntwo
          + {80 \over 9} \* \ec \* \lntwo
          - {464 \over 3} \* \ecs \* \lntwo
\nonumber\\
&&\mbox{}
          - 232 \* \pi^2 \* \lntwo
          + {16 \over 3} \* \pi^2 \* \ec \* \lntwo
          - {24736 \over 9} \* \lntwos
          + {464 \over 3} \* \ec \* \lntwos
          + {64 \over 3} \* \pi^2 \* \lntwos
          + {12064 \over 9} \* \lntwoc
          \biggr]
\nonumber\\
&&\mbox{}
       + \cas \* \biggl[
          - {1592 \over 9}
          - {20 \over 9} \* \ec
          + {68 \over 3} \* \ecs
          + 14 \* \pi^2
          - {4 \over 3} \* \pi^2 \* \ec
          + {2408 \over 9} \* \lntwo
          - {136 \over 3} \* \ec \* \lntwo
          - {8 \over 3} \* \pi^2 \* \lntwo
\nonumber\\
&&\mbox{}
          - {544 \over 3} \* \lntwos
          \biggr]
          + C^{(2)}_{q{\bar q}}
       + \Lfr  \* \cf \* C^{(1)}_{q{\bar q}} \*  \biggl[
            64
          - 32 \* \ec 
          - 64 \* \lntwo
          \biggr]
\nonumber\\
&&\mbox{}
       + \Lfr  \* \cf \* \nf \*  \biggl[
          - {544 \over 9} 
          + {80 \over 9} \* \ec
          + {16 \over 3} \* \ecs
          + {8 \over 3} \* \pi^2
          + {736 \over 9} \* \lntwo
          - {32 \over 3} \* \ec \* \lntwo
          - {128 \over 3} \* \lntwos
          \biggr]
\nonumber\\
&&\mbox{}
       + \Lfr  \* \cfs \*  \biggl[
            1024 
          - 896  \* \z3
          + 512  \* \ec
          + 128  \* \ecs
          - 64  \* \ect
          - 64  \* \pi^2
          - 32  \* \pi^2 \* \ec
          - 1536  \* \lntwo
\nonumber\\
&&\mbox{}
          - 256  \* \ec \* \lntwo
          + 128  \* \ecs \* \lntwo
          + 192  \* \pi^2 \* \lntwo
          + 1664  \* \lntwos
          - 64  \* \ec \* \lntwos
          - 1152  \* \lntwoc
          \biggr]
\nonumber\\
&&\mbox{}
       + \Lfr  \* \ca \* \nf \* \biggl[
          - {16 \over 3}
          + {8 \over 3} \* \ec
          + {16 \over 3} \* \lntwo
          \biggr]
       + \Lfr  \* \ca \* \cf \* \biggl[
            {3184 \over 9}
          + {616 \over 9} \* \ec
          - {184 \over 3} \* \ecs
\nonumber\\
&&\mbox{}
          - 36 \* \pi^2
          + {8 \over 3} \* \pi^2 \* \ec
          - {4816 \over 9} \* \lntwo
          + {272 \over 3} \* \ec \* \lntwo
          + {16 \over 3} \* \pi^2 \* \lntwo
          + {1280 \over 3} \* \lntwos
          \biggr]
\nonumber\\
&&\mbox{}
       + \Lfr  \* \cas \* \biggl[
            {88 \over 3} 
          - {44 \over 3} \* \ec
          - {88 \over 3} \* \lntwo
          \biggr]
       + \Lfrs  \* \cf \* \nf \*  \biggl[
            {16 \over 3}
          - {8 \over 3} \* \ec
          - {16 \over 3} \* \lntwo
          \biggr]
\nonumber\\
&&\mbox{}
       + \Lfrs  \* \cfs \*  \biggl[
          - 128 \* \ec
          + 32 \* \ecs
          + 16 \* \pi^2
          - 128 \* \lntwos
          \biggr]
\nonumber\\
&&\mbox{}
       + \Lfrs  \* \ca \* \cf \*  \biggl[
          - {88 \over 3}
          + {44 \over 3} \* \ec
          + {88 \over 3} \* \lntwo
          \biggr]
          \Biggr\}
\, ,
\\
  \label{eq:g0gg1}
  g^0_{gg,\, 1} &=&
  1 
       + a_s  \*  \Biggl\{
       \ca \* \biggl[
          - 64 
          + 8 \* \ecs
          + 4 \* \pi^2
          - 32 \* \lntwo
          + 8 \* \lntwos
       \biggr]
          + 4 \* C^{(1)}_{gg}
       + \Lqr  \*  \ca \* \biggl[
           16
          - 8 \* \ec
          \biggr]
          \Biggr\}
\nonumber\\
&&\mbox{}
       + a_s^2  \*  \Biggl\{
       \ca \* C^{(1)}_{gg} \* \biggl[
          - 256
          + 32 \* \ecs
          + 16 \* \pi^2
          + 384 \* \lntwo
          - 64 \* \ec \* \lntwo
          - 256 \* \lntwos
          \biggr]
       + \ca \* \nf \* \biggl[
            {6976 \over 27}
\nonumber\\
&&\mbox{}
          - {448 \over 9} \* \z3
          - {224 \over 27} \* \ec
          - {80 \over 9} \* \ecs
          - {32 \over 9} \* \ect
          - {136 \over 9} \* \pi^2
          - {10240 \over 27} \* \lntwo
          + {160 \over 9} \* \ec \* \lntwo
          + {32 \over 3} \* \ecs \* \lntwo
\nonumber\\
&&\mbox{}
          + 16 \* \pi^2 \* \lntwo
          + {2368 \over 9} \* \lntwos
          - {32 \over 3} \* \ec \* \lntwos
          - {832 \over 9} \* \lntwoc
          \biggr]
       + \cas \* \biggl[
          - {262624 \over 27}
          + {35728 \over 9} \* \z3
\nonumber\\
&&\mbox{}
          + {1616 \over 27} \* \ec
          - 56 \* \ec \* \z3
          - {4072 \over 9} \* \ecs
          + {176 \over 9} \* \ect
          + 32 \* \ecf
          + {3292 \over 9} \* \pi^2
          + {88 \over 3} \* \pi^2 \* \ecs
          + {68 \over 3} \* \pi^4
\nonumber\\
&&\mbox{}
          + {392320 \over 27} \* \lntwo
          - 5488 \* \z3 \* \lntwo
          + {8144 \over 9} \* \ec \* \lntwo
          - {176 \over 3} \* \ecs \* \lntwo
          - 128 \* \ect \* \lntwo
          - 888 \* \pi^2 \* \lntwo
\nonumber\\
&&\mbox{}
          - {176 \over 3} \* \pi^2 \* \ec \* \lntwo
          - {101344 \over 9} \* \lntwos
          + {176 \over 3} \* \ec \* \lntwos
          + 192 \* \lntwos \* \ecs
          + {1888 \over 3} \* \pi^2 \* \lntwos
          + {66784 \over 9} \* \lntwoc
\nonumber\\
&&\mbox{}
          - 128 \* \ec \* \lntwoc
          - 2560 \* \lntwof
          \biggr]
          + C^{(2)}_{gg}
       + \Lfr  \* \ca \* C^{(1)}_{gg} \*  \biggl[
            64 
          - 32 \* \ec 
          - 64 \* \lntwo 
          \biggr]
\nonumber\\
&&\mbox{}
       + \Lfr  \* \ca \* \nf \* \biggl[
          - {544 \over 9} 
          + {80 \over 9} \* \ec
          + {16 \over 3} \* \ec^2
          + {8 \over 3} \* \pi^2
          + {736 \over 9}  \* \lntwo
          - {32 \over 3} \* \ec \* \lntwo
          - {128 \over 3} \* \lntwos
          \biggr]
\nonumber\\
&&\mbox{}
       + \Lfr  \* \cas \* \biggl[
            {12400 \over 9} 
          - 896  \* \z3
          + {4072 \over 9}  \* \ec
          + {296 \over 3}  \* \ec^2
          - 64  \* \ec^3
          - 84  \* \pi^2
          - {88 \over 3}  \* \pi^2 \* \ec
\nonumber\\
&&\mbox{}
          - {18064 \over 9}  \* \lntwo
          - {592 \over 3} \* \ec \* \lntwo
          + 128  \* \ec^2 \* \lntwo
          + {592 \over 3} \* \pi^2 \* \lntwo
          + {5696 \over 3}  \* \lntwos
          - 64  \* \ec \* \lntwos
\nonumber\\
&&\mbox{}
          - 1152  \* \lntwoc
          \biggr]
       + \Lfrs  \* \ca \* \nf \* \biggl[
            {16 \over 3} 
          - {8 \over 3} \* \ec
          - {16 \over 3} \* \lntwo
          \biggr]
       + \Lfrs  \* \cas \* \biggl[
          - {88 \over 3} 
          - {340 \over 3}  \* \ec
\nonumber\\
&&\mbox{}
          + 32  \* \ec^2
          + 16  \* \pi^2
          + {88 \over 3}  \* \lntwo
          - 128  \* \lntwos
          \biggr]
          \Biggr\}
\, ,
\\
  \label{eq:g0res}
  g^0_{gg,\, 8} &=&   g^0_{gg,\, 1}
       + a_s  \*  \Biggl\{
       \ca \* \biggl[
          - 8
          + 4 \* \ec
          - 4 \* \lntwo
          \biggr]
          \Biggr\}
\nonumber\\
&&\mbox{}
       + a_s^2  \*  \Biggl\{
       \ca \* C^{(1)}_{gg} \* \biggl[
          - 32
          + 16 \* \ec
          + 32 \* \lntwo
          \biggr]
       + \ca \* \nf \* \biggl[
            {272 \over 9} 
          - {40 \over 9} \* \ec
          - {8 \over 3} \* \ecs
          - {4 \over 3} \* \pi^2
          - {368 \over 9} \* \lntwo
\nonumber\\
&&\mbox{}
          + {16 \over 3} \* \ec \* \lntwo
          + {64 \over 3} \* \lntwos
          \biggr]
       + \cas \* \biggl[
          - {6200 \over 9}
          + 448 \* \z3
          - {2324 \over 9} \* \ec
          - {124 \over 3} \* \ec^2
          + 32 \* \ec^3
          + 46 \* \pi^2
\nonumber\\
&&\mbox{}
          + {44 \over 3} \* \pi^2 \* \ec
          + {11624 \over 9} \* \lntwo
          + {248 \over 3}  \* \ec \* \lntwo
          - 96  \* \ec^2 \* \lntwo
          - {344 \over 3} \* \pi^2 \* \lntwo
          - {4192 \over 3} \* \lntwos
          + 96 \* \ec \* \lntwos
\nonumber\\
&&\mbox{}
          + 832 \* \lntwoc
          \biggr]
       + \Lfr  \* \ca \* \nf \* \biggl[
          - {16 \over 3} 
          + {8 \over 3}  \* \ec
          + {16 \over 3}  \* \lntwo
          \biggr]
       + \Lfr  \* \cas \* \biggl[
            {88 \over 3}
          + {340 \over 3} \* \ec
\nonumber\\
&&\mbox{}
          - 32 \* \ec^2
          - 16 \* \pi^2
          - {280 \over 3} \* \lntwo
          + 32 \* \ec \* \lntwo
          + 192 \* \lntwos
          )
          \Biggr\}
\, ,
\end{eqnarray}
where $C^{(1)}_{q{\bar q}} = 36 \pi a_q^0 + (\nf-4) (2/3 \lntwo-5/9)$ and 
$C^{(1)}_{gg} = 768/7 \pi a_g^0$ with the numerical constants $a_q^0 = 0.180899$ and 
$a_g^0 = 0.108068$ being reported in Table~1 of Ref.~\cite{Nason:1988xz}.
The presently unknown two-loop constants are denoted $C^{(2)}_{q{\bar q}}$ 
and $C^{(2)}_{gg}$.

At first and second order in $\alpha_s$ the Coulomb corrections have to be
added to the cross section. In the limit $\rho \to 1$ (that is $\beta \to 0$) 
they read~\cite{Bernreuther:2004ih,Czarnecki:1997vz}
for the color-singlet final state
\begin{eqnarray}
  \label{eq:sigmaCoul}
  \hat{\sigma}^{(1),\,\rm c}_{ij,\, 1} &=&
  \hat{\sigma}^{(0)}_{ij,\, 1} \, 
  2 \* \cf \* {\pi^2 \over \beta}
\, ,
\\
  \hat{\sigma}^{(2),\,\rm c}_{ij,\, 1} &=&
  \hat{\sigma}^{(0)}_{ij,\, 1} \, 
  2\* \cf \* \Biggl\{
  \left(
          {31 \over 9}\*\ca 
        - 16\*\cf 
        - {10 \over 9}\*\nf 
        - 2\*\b0 \* \lntwo 
        - 2\*\b0 \* \lnbeta
        \right) \* {\pi^2 \over \beta}
+
  {2 \over 3} \* \cf \* {\pi^4 \over \beta^2}
  \Biggr\}
\, .
\qquad
\end{eqnarray}
From these expressions the octet results for $\hat{\sigma}^{(1),\,\rm c}_{ij,\, 8}$ and 
$\hat{\sigma}^{(2),\,\rm c}_{ij,\, 8}$ are obtained by the replacement of the color factor $2\cf \to (2\cf - \ca)$ 
consistent with results from potential non-relativistic QCD (e.g. \cite{Czarnecki:1997vz,Pineda:2006ri}).

Next we give the Mellin transforms of powers of logarithms in $\beta$
according to Eq.~(\ref{eq:mellindef}). Recall that $\rho=4\mt^2/s$ and $\beta=\sqrt{1-4\mt^2/s}=\sqrt{1-\rho}$.
In the limit $\rho \to 1$ as needed in the fixed order expansions and accurate up to power suppressed terms in $N$ we have 
\begin{eqnarray}
  \label{eq:lnmellin}
  \int\limits_{0}^{1}\,d\rho\, \rho^{N}\, \beta \ln^4 \beta &=& 
\left( {1 \over 16} \* \lnNf + \left\{ - {1 \over 2} + {1 \over 2} \* \lntwo \right\} \* \lnNc 
  + \left\{ -3 \* \lntwo + {3 \over 16} \* \pi^2 + {3 \over 2} \* \lntwos \right\} \* \lnNs 
\right.
\nonumber\\& &
\left.
  + \left\{ - {3 \over 4} \* \pi^2 - 6 \* \lntwos + {3 \over 4} \* \pi^2 \* \lntwo + {7 \over 2} \* \z3 + 2 \* \lntwoc \right\} \* \lnN 
  - 7 \* \z3 - 4 \* \lntwoc + {7 \over 64} \* \pi^4
\right.
\nonumber\\& &
\left.
 + 7 \* \z3 \* \lntwo + \lntwof + {3 \over 4} \* \pi^2 \* \lntwos - {3 \over 2} \* \pi^2 \* \lntwo 
\right) \*
{\sqrt{\pi} \over 2}\, {1 \over N^{3/2}}\, \left( 1 + {\cal O}(1/N) \right)\, ,
\nonumber \\
  \int\limits_{0}^{1}\,d\rho\, \rho^{N}\, \beta \ln^3 \beta &=&
\left( - {1 \over 8} \* \lnNc + \left\{ - {3 \over 4} \* \lntwo+ {3 \over 4} \right\} \* \lnNs 
  + \left\{ 3 \* \lntwo - {3 \over 2} \* \lntwos - {3 \over 16} \* \pi^2 \right\} \* \lnN 
\right.
\nonumber\\& &
\left.
  - {7 \over 4} \* \z3- \lntwoc + 3 \* \lntwos + {3 \over 8} \* \pi^2 - {3 \over 8} \* \pi^2 \* \lntwo \right) \*
{\sqrt{\pi} \over 2}\, {1 \over N^{3/2}}\, \left( 1 + {\cal O}(1/N) \right)\, ,
\nonumber \\
  \int\limits_{0}^{1}\,d\rho\, \rho^{N}\, \beta \ln^2 \beta &=&
\left( {1 \over 4} \* \lnNs+ \{\lntwo-1\} \* \lnN + {1 \over 8} \* \pi^2 - 2 \* \lntwo + \lntwos \right) \*
{\sqrt{\pi} \over 2}\, {1 \over N^{3/2}}\, \left( 1 + {\cal O}(1/N) \right)\, ,
\nonumber \\
  \int\limits_{0}^{1}\,d\rho\, \rho^{N}\, \beta \ln \beta &=&
\left( - {1 \over 2} \* \lnN + 1 - \lntwo \right) \*
{\sqrt{\pi} \over 2}\, {1 \over N^{3/2}}\, \left( 1 + {\cal O}(1/N) \right)\, ,
\nonumber \\
  \int\limits_{0}^{1}\,d\rho\, \rho^{N}\,\ln^2 \beta &=&
\left( {1 \over 4} \* \lnNs + {1 \over 24} \* \pi^2 \right) \*
{1 \over N}\, \left( 1 + {\cal O}(1/N) \right)\, ,
\nonumber \\
  \int\limits_{0}^{1}\,d\rho\, \rho^{N}\,\ln \beta &=&
- {1 \over 2} \* \lnN \* {1 \over N}\, \left( 1 + {\cal O}(1/N) \right)\, ,
\end{eqnarray}
where $\tilde{N} = N \exp(\gamma_e)$.

Finally, we present analytical results in the \MSbar-scheme 
for the threshold expansion of the inclusive partonic cross sections
$\hat{\sigma}_{ij \to {t\bar t}} (s,m^2,\mu^2)$. 
The inverse powers of $\beta$ originate from the Coulomb corrections and 
$n_f$ denotes the number of effectively massless quark flavors.
We set $\mu = \mt$ and find for the $q\bar{q}$ channel 
\begin{eqnarray}
\label{eq:sqq10}
  \hat{\sigma}^{(1)}_{q\bar{q}} &=& 
  \hat{\sigma}^{(0)}_{q\bar{q}} 
  \Biggl\{
         32 \* \cf \* \lnbetas
     + \left( 
         96 \* \cf \* \lntwo 
       - 64 \* \cf 
       - 8 \* \ca 
     \right) \* \lnbeta
     + \left( 
         2 \* \cf 
       - \ca 
     \right) \* {\pi^2  \over \beta}
\nonumber\\
&&
       + 4 \* C^{(1)}_{q{\bar q}} 
       - 96 \* \cf \* \lntwo 
       + 72 \* \cf \* \lntwos 
       - 12 \* \ca \* \lntwo
    \Biggr\}
\, ,
\\ 
\label{eq:sqq20}
  \hat{\sigma}^{(2)}_{q\bar{q}} &=& 
  \hat{\sigma}^{(0)}_{q\bar{q}} 
  \Biggl\{  
         512 \* \cfs \* \lnbetaf
     + \biggl( 
         {256 \over 9} \* \cf \* \nf 
       - 2048 \* \cfs 
       + 3072 \* \cfs \* \lntwo 
       - {3712 \over 9} \* \ca \* \cf 
     \biggr) \* \lnbetac
\nonumber \\
&&
     + \biggl( 
         128 \* \cf \* C^{(1)}_{q{\bar q}} 
       - {1088 \over 9} \* \cf \* \nf 
       + 128 \* \cf \* \nf \* \lntwo 
       + 4096 \* \cfs 
       - 256 \* \cfs \* \pi^2 
       - 9216 \* \cfs \* \lntwo 
\nonumber \\
&&
       + 6912 \* \cfs \* \lntwos 
       - {32 \over 3} \* \ca \* \nf 
       + {10976 \over 9} \* \ca \* \cf 
       - {32 \over 3} \* \ca \* \cf \* \pi^2 
       - 1856 \* \ca \* \cf \* \lntwo 
\nonumber \\
&&
       + {272 \over 3} \* \cas 
     \biggr) \* \lnbetas
     + \left( 
         64 \* \cfs
       - 32 \* \ca \* \cf
     \right) \* \lnbetas \* {\pi^2 \over \beta}
     + \biggl(  
         384 \* \cf \* C^{(1)}_{q{\bar q}} \* \lntwo
       - 256 \* \cf \* C^{(1)}_{q{\bar q}} 
\nonumber \\
&&
       + {6976 \over 27} \* \cf \* \nf 
       - {32 \over 3} \* \cf \* \nf \* \pi^2 
       - {1088 \over 3} \* \cf \* \nf \* \lntwo 
       + 192 \* \cf \* \nf \* \lntwos 
       - 8192 \* \cfs 
       + 3584 \* \cfs \* \z3
\nonumber \\
&&
       + 512 \* \cfs \* \pi^2 
       + 12288 \* \cfs \* \lntwo 
       - 768 \* \cfs \* \pi^2 \* \lntwo 
       - 13824 \* \cfs \* \lntwos 
       + 6912 \* \cfs \* \lntwoc 
\nonumber \\
&&
       - 32 \* \ca \* C^{(1)}_{q{\bar q}} 
       + {272 \over 9} \* \ca \* \nf 
       - 32 \* \ca \* \nf \* \lntwo 
       - {55264 \over 27} \* \ca \* \cf 
       + 112 \* \ca \* \cf \* \z3 
       + 144 \* \ca \* \cf \* \pi^2 
\nonumber \\
&&
       + {10976 \over 3} \* \ca \* \cf \* \lntwo 
       - 32 \* \ca \* \cf \* \pi^2 \* \lntwo 
       - 2784 \* \ca \* \cf \* \lntwos 
       - {1592 \over 9} \* \cas 
       + {8 \over 3} \* \cas \* \pi^2 
\nonumber \\
&&
       + 272 \* \cas \* \lntwo 
     \biggr) \* \lnbeta
     + \biggl( 
         {8 \over 3} \* \cf \* \nf 
       + 64 \* \cfs \* \lntwo 
       - {4 \over 3} \* \ca \* \nf
       - {92 \over 3} \* \ca \* \cf
       - 32 \* \ca \* \cf \* \lntwo
\nonumber \\
&& 
       + {46 \over 3} \* \cas 
     \biggr) \* \lnbeta \* {\pi^2 \over \beta}
     + {4 \over 3} \* \biggl( 
         \cf
       - {1 \over 2} \* \ca 
     \biggr)^2 \* {\pi^4 \over \beta^2}
     + \biggl(  
         {8 \over 3} \* \cf \* \nf \* \lntwo 
       - {20 \over 9} \* \cf \* \nf 
       - 32 \* \cfs 
       + {10 \over 9} \* \ca \* \nf 
\nonumber \\
&&
       - {4 \over 3} \* \ca \* \nf \* \lntwo 
       + {350 \over 9} \* \ca \* \cf 
       - {44 \over 3} \* \ca \* \cf \* \lntwo 
       - {103 \over 9} \* \cas 
       + {22 \over 3} \* \cas \* \lntwo 
     \biggr) \* {\pi^2 \over \beta}
       + C^{(2)}_{q{\bar q}}
    \Biggr\}
\, ,
\end{eqnarray}
where $C^{(1)}_{q{\bar q}}$ has been given below Eq.~(\ref{eq:g0res}) and 
$C^{(2)}_{q{\bar q}}$ is presently unknown.
For the $gg$ channel at scale $\mu = \mt$, we find
\begin{eqnarray}
\label{eq:sgg10}
  \hat{\sigma}^{(1)}_{gg} &=& 
  \hat{\sigma}^{(0)}_{gg} 
  \Biggl\{
         32 \* \ca \* \lnbetas
       + \left(
            96 \* \ca \* \lntwo
          - 72 \* \ca
          + 16 \* \ca  \* {1 \over \nc^2 - 2}
          \right) \* \lnbeta
       + \biggl(
            2 \* \cf
          - \ca
\nonumber\\
&& 
         + 2 \* \ca \* {1 \over \nc^2 - 2}
          \biggr) \* {\pi^2 \over \beta}
       + 4 \* C^{(1)}_{gg}
          - 108 \* \ca \* \lntwo
          + 72 \* \ca \* \lntwos
          + 24 \* \ca \* \lntwo  \* {1 \over \nc^2 - 2}
    \Biggr\}
\, ,
\\ 
\label{eq:sgg20}
  \hat{\sigma}^{(2)}_{gg} &=& 
  \hat{\sigma}^{(0)}_{gg} 
  \Biggl\{  
          512 \* \cas \* \lnbetaf
       + \left(
            512
          + {256 \over 9} \* \ca \* \nf
          - {22144 \over 9} \* \cas
          + 3072 \* \cas \* \lntwo
          + 1024 \* {1 \over \nc^2 - 2}
       \right) \* \lnbetac
\nonumber\\
&& 
       + \biggl(
            2304 \* \lntwo
          - {3616 \over 3}
          + 128 \* \ca \* C^{(1)}_{gg}
          - {1184 \over 9} \* \ca \* \nf
          + 128 \* \ca \* \nf \* \lntwo
          + {48656 \over 9} \* \cas
          - {800 \over 3} \* \cas \* \pi^2
\nonumber\\
&& 
          - 11072 \* \cas \* \lntwo
          + 6912 \* \cas \* \lntwos
          + \biggl[
            4608 \* \lntwo
          - {7232 \over 3}
          + {64 \over 3} \* \ca \* \nf
          \biggr] \* {1 \over \nc^2 - 2} 
       \biggr) \* \lnbetas
\nonumber\\
&& 
       + \biggl(
            64
          + 64 \* \ca \* \cf
          - 32 \* \cas
          + 128 \* {1 \over \nc^2 - 2}
       \biggr) \* \lnbetas \* {\pi^2 \over \beta}
       + \biggl(
            {12400 \over 9}
          - {400 \over 3} \* \pi^2
          - 3616 \* \lntwo
\nonumber\\
&& 
          + 3456 \* \lntwos
          - 288 \* \ca \* C^{(1)}_{gg}
          + 384 \* \ca \* C^{(1)}_{gg} \* \lntwo
          + {7792 \over 27} \* \ca \* \nf
          - {32 \over 3} \* \ca \* \nf \* \pi^2
          - {1184 \over 3} \* \ca \* \nf \* \lntwo
\nonumber\\
&& 
          + 192 \* \ca \* \nf \* \lntwos
          - {281224 \over 27} \* \cas
          + 3696 \* \cas \* \z3
          + {1976 \over 3} \* \cas \* \pi^2
          + {48656 \over 3} \* \cas \* \lntwo
\nonumber\\
&& 
          - 800 \* \cas \* \pi^2 \* \lntwo
          - 16608 \* \cas \* \lntwos
          + 6912 \* \cas \* \lntwoc
          + \biggl[
            {24800 \over 9}
          - {800 \over 3} \* \pi^2
          - 7232 \* \lntwo
\nonumber\\
&& 
          + 6912 \* \lntwos
          + 64 \* \ca \* C^{(1)}_{gg}
          - {544 \over 9} \* \ca \* \nf
          + 64 \* \ca \* \nf \* \lntwo
          \biggr] \* {1 \over \nc^2 - 2}
       \biggr) \* \lnbeta
       + \biggl(
            64 \* \lntwo
          - {44 \over 3}
\nonumber\\
&& 
          + {8 \over 3} \* \cf \* \nf
          - {4 \over 3} \* \ca \* \nf
          - {92 \over 3} \* \ca \* \cf
          + 64 \* \ca \* \cf \* \lntwo
          + {46 \over 3} \* \cas
          - 32 \* \cas \* \lntwo
          +\biggl[
            128 \* \lntwo
\nonumber\\
&& 
          - {136 \over 3}
          + {8 \over 3} \* \ca \* \nf
          \biggr] \* {1 \over \nc^2 - 2}
       \biggr) \* \lnbeta \* {\pi^2 \over \beta}
       + \biggl[ {4 \over 3} \* \biggl(
            \cf
          - {1 \over 2} \* \ca
          \biggr)^2
          + {2 \over 3}
       \biggr] \* {\pi^4 \over \beta^2}
       + \biggl(
          - {82 \over 9}
          - {44 \over 3} \* \lntwo
\nonumber\\
&& 
          - {20 \over 9} \* \cf \* \nf
          + {8 \over 3} \* \cf \* \nf \* \lntwo
          - 32 \* \cfs
          + {10 \over 9} \* \ca \* \nf
          - {4 \over 3} \* \ca \* \nf \* \lntwo
          + {350 \over 9} \* \ca \* \cf
          - {44 \over 3} \* \ca \* \cf \* \lntwo
\nonumber\\
&& 
          - {103 \over 9} \* \cas
          + {22 \over 3} \* \cas \* \lntwo
          + \biggl[
            {124 \over 9}
          - {88 \over 3} \* \lntwo
          - {20 \over 9} \* \ca \* \nf
          + {8 \over 3} \* \ca \* \nf \* \lntwo
          \biggr] \* {1 \over \nc^2 - 2}
       \biggr) \* {\pi^2 \over \beta}
\nonumber\\
&& 
       + C^{(2)}_{gg}
    \Biggr\}
\, ,
\end{eqnarray}
and $C^{(1)}_{gg}$ has again been given below Eq.~(\ref{eq:g0res}) while 
$C^{(2)}_{gg}$ is the yet uncalculated two-loop constant. 

\newpage
\section{Detailed results for specific PDFs}
\begin{table}[htbp]
\begin{center}
\leavevmode\small
\renewcommand{\arraystretch}{1.4}
\begin{tabular}{c|ccc|c|ccc} 
 pdf set&
\multicolumn{1}{c}{$\sigmaNLO$ }&
\multicolumn{1}{c}{$\sigmaRES$}&
\multicolumn{1}{c|}{$\sigmaNNLO$}&
 pdf set&
\multicolumn{1}{c}{$\sigmaNLO$ }&
\multicolumn{1}{c}{$\sigmaRES$}&
\multicolumn{1}{c}{$\sigmaNNLO$}
\\ \hline 
0 & 925$^{ -108}_{+112}$ & 932$^{ -94}_{+105}$ & 969$^{ -13}_{ -39}$&
\\ 
\hline
1 & 920$^{ -108}_{+111}$ & 927$^{ -94}_{+105}$ & 964$^{ -12}_{ -39}$ & 2 & 929$^{ -109}_{+112}$ & 937$^{ -95}_{+106}$ & 974$^{ -13}_{ -39}$\\ 
3 & 924$^{ -108}_{+111}$ & 931$^{ -94}_{+105}$ & 968$^{ -12}_{ -39}$ & 4 & 925$^{ -109}_{+112}$ & 933$^{ -95}_{+106}$ & 970$^{ -13}_{ -39}$\\ 
5 & 928$^{ -109}_{+112}$ & 935$^{ -95}_{+106}$ & 972$^{ -13}_{ -39}$ & 6 & 922$^{ -108}_{+111}$ & 929$^{ -94}_{+105}$ & 966$^{ -12}_{ -39}$\\ 
7 & 925$^{ -108}_{+112}$ & 932$^{ -94}_{+105}$ & 969$^{ -13}_{ -39}$ & 8 & 924$^{ -108}_{+112}$ & 931$^{ -94}_{+105}$ & 969$^{ -13}_{ -39}$\\ 
9 & 926$^{ -109}_{+112}$ & 933$^{ -94}_{+105}$ & 970$^{ -13}_{ -39}$ & 10 & 923$^{ -108}_{+112}$ & 931$^{ -94}_{+105}$ & 968$^{ -12}_{ -39}$\\ 
11 & 927$^{ -108}_{+112}$ & 934$^{ -94}_{+106}$ & 972$^{ -12}_{ -39}$ & 12 & 922$^{ -108}_{+112}$ & 929$^{ -94}_{+105}$ & 966$^{ -13}_{ -39}$\\ 
13 & 926$^{ -108}_{+112}$ & 934$^{ -94}_{+105}$ & 971$^{ -13}_{ -39}$ & 14 & 923$^{ -108}_{+112}$ & 930$^{ -94}_{+105}$ & 967$^{ -13}_{ -39}$\\ 
15 & 929$^{ -109}_{+112}$ & 936$^{ -95}_{+106}$ & 973$^{ -13}_{ -39}$ & 16 & 920$^{ -108}_{+111}$ & 927$^{ -94}_{+105}$ & 965$^{ -12}_{ -39}$\\ 
17 & 916$^{ -108}_{+111}$ & 924$^{ -94}_{+104}$ & 961$^{ -13}_{ -39}$ & 18 & 932$^{ -109}_{+112}$ & 939$^{ -94}_{+106}$ & 976$^{ -12}_{ -40}$\\ 
19 & 922$^{ -108}_{+111}$ & 929$^{ -94}_{+105}$ & 966$^{ -12}_{ -39}$ & 20 & 926$^{ -109}_{+112}$ & 933$^{ -95}_{+106}$ & 971$^{ -13}_{ -39}$\\ 
21 & 925$^{ -108}_{+112}$ & 932$^{ -94}_{+105}$ & 970$^{ -13}_{ -39}$ & 22 & 924$^{ -108}_{+112}$ & 931$^{ -94}_{+105}$ & 968$^{ -13}_{ -39}$\\ 
23 & 924$^{ -109}_{+112}$ & 931$^{ -94}_{+105}$ & 969$^{ -13}_{ -39}$ & 24 & 925$^{ -108}_{+112}$ & 933$^{ -94}_{+105}$ & 970$^{ -12}_{ -39}$\\ 
25 & 924$^{ -108}_{+112}$ & 931$^{ -94}_{+105}$ & 969$^{ -12}_{ -39}$ & 26 & 925$^{ -108}_{+112}$ & 932$^{ -94}_{+105}$ & 969$^{ -13}_{ -39}$\\ 
27 & 926$^{ -108}_{+112}$ & 933$^{ -94}_{+105}$ & 970$^{ -13}_{ -39}$ & 28 & 925$^{ -109}_{+112}$ & 932$^{ -95}_{+105}$ & 970$^{ -13}_{ -39}$\\ 
29 & 924$^{ -108}_{+111}$ & 931$^{ -94}_{+105}$ & 968$^{ -12}_{ -39}$ & 30 & 924$^{ -108}_{+112}$ & 931$^{ -94}_{+105}$ & 969$^{ -12}_{ -39}$\\ 
\hline
\end{tabular}
\renewcommand{\arraystretch}{1.}
\end{center}
\caption{ \small
The total cross section for $\mu = \mt$ at LHC
 for $\mt$ = 171 GeV and the full set of predictions
from the MRST-2006 NNLO PDF set~\cite{Martin:2007bv}. All rates are in pb. We denote by \sigmaNLO\ \cite{Nason:1988xz,Beenakker:1989bq} the 
NLO QCD prediction by \sigmaRES\ 
the result of NLL threshold resummation \cite{Bonciani:1998vc}
and by \sigmaNNLO\ the NNLO QCD prediction based on soft
gluon approximation and exact two-loop scale dependence 
\cite{Kidonakis:2001nj}. The upper and lower indices denote
the shifts towards $\mu = 2\mt$ and $\mu=\mt/2$.}
 \end{table}
\begin{table}[htbp]
\begin{center}
\leavevmode\small
\renewcommand{\arraystretch}{1.4}
\begin{tabular}{c|ccc|c|ccc} 
 pdf set&
\multicolumn{1}{c}{$\sigmaNLO$ }&
\multicolumn{1}{c}{$\sigmaRES$}&
\multicolumn{1}{c|}{$\sigmaNNLO$}&
 pdf set&
\multicolumn{1}{c}{$\sigmaNLO$ }&
\multicolumn{1}{c}{$\sigmaRES$}&
\multicolumn{1}{c}{$\sigmaNNLO$}
\\ \hline 
0 & 875$^{ -101}_{+102}$ & 882$^{ -87}_{+95}$ & 918$^{ -9}_{ -39}$&
\\ 
\hline
1 & 877$^{ -101}_{+102}$ & 884$^{ -87}_{+96}$ & 920$^{ -9}_{ -39}$ & 2 & 873$^{ -101}_{+102}$ & 880$^{ -87}_{+95}$ & 916$^{ -9}_{ -39}$\\ 
3 & 878$^{ -101}_{+102}$ & 885$^{ -88}_{+96}$ & 921$^{ -9}_{ -39}$ & 4 & 872$^{ -101}_{+101}$ & 879$^{ -87}_{+95}$ & 915$^{ -9}_{ -39}$\\ 
5 & 877$^{ -101}_{+102}$ & 884$^{ -88}_{+96}$ & 921$^{ -9}_{ -39}$ & 6 & 872$^{ -100}_{+101}$ & 879$^{ -87}_{+95}$ & 915$^{ -9}_{ -39}$\\ 
7 & 883$^{ -103}_{+103}$ & 890$^{ -89}_{+97}$ & 926$^{ -10}_{ -40}$ & 8 & 867$^{ -99}_{+100}$ & 874$^{ -86}_{+94}$ & 909$^{ -9}_{ -39}$\\ 
9 & 877$^{ -101}_{+102}$ & 884$^{ -87}_{+96}$ & 920$^{ -9}_{ -39}$ & 10 & 872$^{ -101}_{+102}$ & 879$^{ -87}_{+95}$ & 915$^{ -9}_{ -39}$\\ 
11 & 899$^{ -103}_{+105}$ & 906$^{ -89}_{+98}$ & 943$^{ -9.6}_{ -40}$ & 12 & 852$^{ -99}_{+99}$ & 858$^{ -85}_{+93}$ & 894$^{ -9}_{ -38}$\\ 
13 & 876$^{ -101}_{+102}$ & 883$^{ -88}_{+96}$ & 919$^{ -9}_{ -39}$ & 14 & 874$^{ -101}_{+101}$ & 881$^{ -87}_{+95}$ & 917$^{ -9}_{ -39}$\\ 
15 & 869$^{ -101}_{+102}$ & 876$^{ -87}_{+95}$ & 912$^{ -9}_{ -39}$ & 16 & 880$^{ -101}_{+102}$ & 887$^{ -87}_{+96}$ & 923$^{ -9}_{ -39}$\\ 
17 & 880$^{ -102}_{+103}$ & 888$^{ -88}_{+96}$ & 924$^{ -9.5}_{ -40}$ & 18 & 869$^{ -100}_{+101}$ & 876$^{ -86}_{+95}$ & 912$^{ -9}_{ -39}$\\ 
19 & 878$^{ -101}_{+102}$ & 885$^{ -88}_{+96}$ & 921$^{ -9}_{ -39}$ & 20 & 871$^{ -101}_{+101}$ & 878$^{ -87}_{+95}$ & 914$^{ -9}_{ -39}$\\ 
21 & 870$^{ -101}_{+101}$ & 877$^{ -87}_{+95}$ & 913$^{ -9}_{ -39}$ & 22 & 880$^{ -101}_{+102}$ & 887$^{ -87}_{+96}$ & 923$^{ -9}_{ -40}$\\ 
23 & 867$^{ -100}_{+101}$ & 874$^{ -87}_{+95}$ & 910$^{ -9}_{ -39}$ & 24 & 885$^{ -102}_{+103}$ & 892$^{ -88}_{+96}$ & 928$^{ -9}_{ -40}$\\ 
25 & 875$^{ -99}_{+101}$ & 882$^{ -86}_{+95}$ & 917$^{ -8}_{ -39}$ & 26 & 875$^{ -102}_{+102}$ & 882$^{ -88}_{+96}$ & 919$^{ -9.7}_{ -39}$\\ 
27 & 874$^{ -101}_{+102}$ & 881$^{ -87}_{+95}$ & 917$^{ -9}_{ -39}$ & 28 & 875$^{ -101}_{+102}$ & 882$^{ -87}_{+96}$ & 918$^{ -9}_{ -39}$\\ 
29 & 876$^{ -101}_{+102}$ & 883$^{ -88}_{+96}$ & 920$^{ -9}_{ -39}$ & 30 & 873$^{ -101}_{+102}$ & 880$^{ -87}_{+95}$ & 916$^{ -9}_{ -39}$\\ 
31 & 868$^{ -99.9}_{+101}$ & 875$^{ -86}_{+95}$ & 911$^{ -9}_{ -39}$ & 32 & 878$^{ -102}_{+102}$ & 886$^{ -88}_{+96}$ & 922$^{ -9.5}_{ -39}$\\ 
33 & 875$^{ -102}_{+102}$ & 882$^{ -88}_{+96}$ & 918$^{ -9.6}_{ -39}$ & 34 & 874$^{ -100}_{+101}$ & 881$^{ -86}_{+95}$ & 917$^{ -9}_{ -39}$\\ 
35 & 873$^{ -101}_{+102}$ & 880$^{ -87}_{+95}$ & 916$^{ -9}_{ -39}$ & 36 & 874$^{ -101}_{+102}$ & 881$^{ -87}_{+95}$ & 917$^{ -9}_{ -39}$\\ 
37 & 875$^{ -101}_{+102}$ & 882$^{ -87}_{+95}$ & 918$^{ -9}_{ -39}$ & 38 & 872$^{ -101}_{+102}$ & 879$^{ -87}_{+95}$ & 916$^{ -9}_{ -39}$\\ 
39 & 874$^{ -101}_{+102}$ & 881$^{ -87}_{+95}$ & 917$^{ -9}_{ -39}$ & 40 & 875$^{ -101}_{+102}$ & 882$^{ -87}_{+96}$ & 918$^{ -9}_{ -39}$\\ 
\hline
\end{tabular}
\renewcommand{\arraystretch}{1.}
\end{center}
\caption{ \small
The total cross section for $\mu = \mt$ at LHC
 for $\mt$ = 171 GeV and the full set of predictions
from the CTEQ6.5 PDF set~\cite{Tung:2006tb}. All rates are in pb. We denote by \sigmaNLO\ \cite{Nason:1988xz,Beenakker:1989bq} the 
NLO QCD prediction by \sigmaRES\ 
the result of NLL threshold resummation \cite{Bonciani:1998vc}
and by \sigmaNNLO\ the NNLO QCD prediction based on soft
gluon approximation and exact two-loop scale dependence 
\cite{Kidonakis:2001nj}. The upper and lower indices denote
the shifts towards $\mu = 2\mt$ and $\mu=\mt/2$.}
 \end{table}

\begin{table}[htbp]
\begin{center}
\leavevmode\small
\renewcommand{\arraystretch}{1.4}
\begin{tabular}{c|ccc|c|ccc} 
 pdf set&
\multicolumn{1}{c}{$\sigmaNLO$ }&
\multicolumn{1}{c}{$\sigmaRES$}&
\multicolumn{1}{c|}{$\sigmaNNLO$}&
 pdf set&
\multicolumn{1}{c}{$\sigmaNLO$ }&
\multicolumn{1}{c}{$\sigmaRES$}&
\multicolumn{1}{c}{$\sigmaNNLO$}
\\ \hline 
0 & 7.35$^{ -0.80}_{+0.38}$ & 7.53$^{ -0.66}_{+0.25}$ & 7.94$^{ -0.28}_{+0.07}$&
\\ 
\hline
1 & 7.48$^{ -0.81}_{+0.38}$ & 7.67$^{ -0.67}_{+0.25}$ & 8.08$^{ -0.29}_{+0.07}$ & 2 & 7.23$^{ -0.79}_{+0.37}$ & 7.41$^{ -0.65}_{+0.25}$ & 7.81$^{ -0.28}_{+0.07}$\\ 
3 & 7.35$^{ -0.80}_{+0.38}$ & 7.54$^{ -0.66}_{+0.25}$ & 7.94$^{ -0.28}_{+0.07}$ & 4 & 7.35$^{ -0.80}_{+0.38}$ & 7.53$^{ -0.66}_{+0.25}$ & 7.94$^{ -0.28}_{+0.07}$\\ 
5 & 7.37$^{ -0.80}_{+0.38}$ & 7.55$^{ -0.66}_{+0.25}$ & 7.96$^{ -0.28}_{+0.07}$ & 6 & 7.33$^{ -0.80}_{+0.38}$ & 7.52$^{ -0.66}_{+0.25}$ & 7.92$^{ -0.28}_{+0.07}$\\ 
7 & 7.22$^{ -0.79}_{+0.38}$ & 7.40$^{ -0.65}_{+0.25}$ & 7.81$^{ -0.28}_{+0.07}$ & 8 & 7.49$^{ -0.81}_{+0.38}$ & 7.67$^{ -0.67}_{+0.25}$ & 8.09$^{ -0.28}_{+0.07}$\\ 
9 & 7.38$^{ -0.81}_{+0.38}$ & 7.56$^{ -0.66}_{+0.26}$ & 7.97$^{ -0.28}_{+0.07}$ & 10 & 7.32$^{ -0.80}_{+0.37}$ & 7.51$^{ -0.66}_{+0.25}$ & 7.91$^{ -0.28}_{+0.07}$\\ 
11 & 7.44$^{ -0.83}_{+0.41}$ & 7.62$^{ -0.68}_{+0.28}$ & 8.04$^{ -0.29}_{+0.06}$ & 12 & 7.29$^{ -0.78}_{+0.36}$ & 7.47$^{ -0.64}_{+0.23}$ & 7.87$^{ -0.27}_{+0.08}$\\ 
13 & 7.34$^{ -0.80}_{+0.38}$ & 7.52$^{ -0.66}_{+0.25}$ & 7.93$^{ -0.28}_{+0.07}$ & 14 & 7.36$^{ -0.80}_{+0.38}$ & 7.54$^{ -0.66}_{+0.25}$ & 7.95$^{ -0.28}_{+0.07}$\\ 
15 & 7.27$^{ -0.79}_{+0.37}$ & 7.45$^{ -0.65}_{+0.24}$ & 7.85$^{ -0.28}_{+0.07}$ & 16 & 7.44$^{ -0.82}_{+0.39}$ & 7.63$^{ -0.67}_{+0.26}$ & 8.04$^{ -0.29}_{+0.07}$\\ 
17 & 7.29$^{ -0.80}_{+0.38}$ & 7.47$^{ -0.66}_{+0.26}$ & 7.87$^{ -0.28}_{+0.07}$ & 18 & 7.40$^{ -0.80}_{+0.38}$ & 7.59$^{ -0.66}_{+0.25}$ & 8.00$^{ -0.28}_{+0.07}$\\ 
19 & 7.28$^{ -0.80}_{+0.38}$ & 7.46$^{ -0.66}_{+0.26}$ & 7.87$^{ -0.28}_{+0.07}$ & 20 & 7.43$^{ -0.80}_{+0.37}$ & 7.61$^{ -0.66}_{+0.25}$ & 8.02$^{ -0.28}_{+0.07}$\\ 
21 & 7.29$^{ -0.79}_{+0.36}$ & 7.47$^{ -0.65}_{+0.24}$ & 7.87$^{ -0.28}_{+0.07}$ & 22 & 7.44$^{ -0.82}_{+0.40}$ & 7.62$^{ -0.68}_{+0.27}$ & 8.04$^{ -0.29}_{+0.06}$\\ 
23 & 7.25$^{ -0.78}_{+0.36}$ & 7.43$^{ -0.64}_{+0.23}$ & 7.82$^{ -0.27}_{+0.08}$ & 24 & 7.50$^{ -0.83}_{+0.41}$ & 7.69$^{ -0.69}_{+0.28}$ & 8.11$^{ -0.29}_{+0.06}$\\ 
25 & 7.71$^{ -0.87}_{+0.45}$ & 7.90$^{ -0.72}_{+0.32}$ & 8.34$^{ -0.30}_{+0.05}$ & 26 & 7.19$^{ -0.77}_{+0.34}$ & 7.37$^{ -0.63}_{+0.22}$ & 7.76$^{ -0.27}_{+0.08}$\\ 
27 & 7.41$^{ -0.81}_{+0.39}$ & 7.59$^{ -0.67}_{+0.26}$ & 8.01$^{ -0.29}_{+0.07}$ & 28 & 7.29$^{ -0.79}_{+0.37}$ & 7.48$^{ -0.65}_{+0.24}$ & 7.88$^{ -0.28}_{+0.07}$\\ 
29 & 7.35$^{ -0.80}_{+0.38}$ & 7.54$^{ -0.66}_{+0.25}$ & 7.94$^{ -0.28}_{+0.07}$ & 30 & 7.34$^{ -0.80}_{+0.38}$ & 7.52$^{ -0.66}_{+0.25}$ & 7.93$^{ -0.28}_{+0.07}$\\ 
31 & 7.42$^{ -0.81}_{+0.39}$ & 7.60$^{ -0.67}_{+0.26}$ & 8.01$^{ -0.29}_{+0.07}$ & 32 & 7.27$^{ -0.79}_{+0.37}$ & 7.45$^{ -0.65}_{+0.24}$ & 7.85$^{ -0.28}_{+0.07}$\\ 
33 & 7.27$^{ -0.79}_{+0.36}$ & 7.46$^{ -0.65}_{+0.24}$ & 7.86$^{ -0.28}_{+0.08}$ & 34 & 7.44$^{ -0.82}_{+0.40}$ & 7.63$^{ -0.68}_{+0.27}$ & 8.04$^{ -0.29}_{+0.06}$\\ 
35 & 7.39$^{ -0.81}_{+0.39}$ & 7.58$^{ -0.67}_{+0.26}$ & 7.99$^{ -0.28}_{+0.07}$ & 36 & 7.32$^{ -0.80}_{+0.37}$ & 7.50$^{ -0.66}_{+0.25}$ & 7.91$^{ -0.28}_{+0.07}$\\ 
37 & 7.37$^{ -0.81}_{+0.38}$ & 7.55$^{ -0.67}_{+0.26}$ & 7.96$^{ -0.28}_{+0.07}$ & 38 & 7.31$^{ -0.79}_{+0.37}$ & 7.49$^{ -0.65}_{+0.25}$ & 7.90$^{ -0.28}_{+0.07}$\\ 
39 & 7.33$^{ -0.80}_{+0.38}$ & 7.52$^{ -0.66}_{+0.25}$ & 7.92$^{ -0.28}_{+0.07}$ & 40 & 7.35$^{ -0.80}_{+0.38}$ & 7.53$^{ -0.66}_{+0.25}$ & 7.94$^{ -0.28}_{+0.07}$\\ 
\hline
\end{tabular}
\renewcommand{\arraystretch}{1.}
\end{center}
\caption{ \small
The total cross section for $\mu = \mt$ at Tevatron
 for $\mt$ = 171 GeV and the full set of predictions
from the CTEQ6.5 PDF set~\cite{Tung:2006tb}. All rates are in pb. We denote by \sigmaNLO\ \cite{Nason:1988xz,Beenakker:1989bq} the 
NLO QCD prediction by \sigmaRES\ 
the result of NLL threshold resummation \cite{Bonciani:1998vc}
and by \sigmaNNLO\ the NNLO QCD prediction based on soft
gluon approximation and exact two-loop scale dependence 
\cite{Kidonakis:2001nj}. The upper and lower indices denote
the shifts towards $\mu = 2\mt$ and $\mu=\mt/2$.}
 \end{table}
\begin{table}[htbp]
\begin{center}
\leavevmode\small
\renewcommand{\arraystretch}{1.4}
\begin{tabular}{c|ccc|c|ccc} 
 pdf set&
\multicolumn{1}{c}{$\sigmaNLO$ }&
\multicolumn{1}{c}{$\sigmaRES$}&
\multicolumn{1}{c|}{$\sigmaNNLO$}&
 pdf set&
\multicolumn{1}{c}{$\sigmaNLO$ }&
\multicolumn{1}{c}{$\sigmaRES$}&
\multicolumn{1}{c}{$\sigmaNNLO$}
\\ \hline 
0 & 7.60$^{ -0.90}_{+0.47}$ & 7.79$^{ -0.75}_{+0.35}$ & 8.24$^{ -0.34}_{+0.08}$&
\\ 
\hline
1 & 7.59$^{ -0.89}_{+0.47}$ & 7.78$^{ -0.75}_{+0.34}$ & 8.23$^{ -0.34}_{+0.08}$ & 2 & 7.62$^{ -0.90}_{+0.48}$ & 7.80$^{ -0.75}_{+0.35}$ & 8.26$^{ -0.34}_{+0.08}$\\ 
3 & 7.63$^{ -0.90}_{+0.48}$ & 7.81$^{ -0.75}_{+0.35}$ & 8.27$^{ -0.34}_{+0.08}$ & 4 & 7.58$^{ -0.89}_{+0.47}$ & 7.77$^{ -0.75}_{+0.34}$ & 8.22$^{ -0.34}_{+0.08}$\\ 
5 & 7.55$^{ -0.89}_{+0.47}$ & 7.73$^{ -0.75}_{+0.34}$ & 8.18$^{ -0.34}_{+0.08}$ & 6 & 7.66$^{ -0.90}_{+0.48}$ & 7.85$^{ -0.75}_{+0.35}$ & 8.30$^{ -0.34}_{+0.08}$\\ 
7 & 7.58$^{ -0.89}_{+0.47}$ & 7.76$^{ -0.75}_{+0.34}$ & 8.21$^{ -0.34}_{+0.08}$ & 8 & 7.63$^{ -0.90}_{+0.47}$ & 7.82$^{ -0.75}_{+0.35}$ & 8.27$^{ -0.34}_{+0.08}$\\ 
9 & 7.60$^{ -0.90}_{+0.47}$ & 7.79$^{ -0.75}_{+0.34}$ & 8.24$^{ -0.34}_{+0.08}$ & 10 & 7.60$^{ -0.90}_{+0.48}$ & 7.79$^{ -0.75}_{+0.35}$ & 8.24$^{ -0.34}_{+0.08}$\\ 
11 & 7.63$^{ -0.90}_{+0.48}$ & 7.82$^{ -0.76}_{+0.35}$ & 8.27$^{ -0.34}_{+0.08}$ & 12 & 7.58$^{ -0.89}_{+0.47}$ & 7.76$^{ -0.74}_{+0.34}$ & 8.21$^{ -0.34}_{+0.09}$\\ 
13 & 7.62$^{ -0.90}_{+0.48}$ & 7.81$^{ -0.75}_{+0.35}$ & 8.26$^{ -0.34}_{+0.08}$ & 14 & 7.60$^{ -0.89}_{+0.47}$ & 7.78$^{ -0.75}_{+0.34}$ & 8.23$^{ -0.34}_{+0.08}$\\ 
15 & 7.65$^{ -0.90}_{+0.48}$ & 7.84$^{ -0.76}_{+0.35}$ & 8.29$^{ -0.34}_{+0.08}$ & 16 & 7.57$^{ -0.89}_{+0.47}$ & 7.75$^{ -0.74}_{+0.34}$ & 8.20$^{ -0.34}_{+0.08}$\\ 
17 & 7.50$^{ -0.87}_{+0.45}$ & 7.68$^{ -0.73}_{+0.32}$ & 8.12$^{ -0.33}_{+0.09}$ & 18 & 7.74$^{ -0.93}_{+0.50}$ & 7.93$^{ -0.78}_{+0.37}$ & 8.40$^{ -0.35}_{+0.08}$\\ 
19 & 7.70$^{ -0.91}_{+0.49}$ & 7.88$^{ -0.77}_{+0.36}$ & 8.34$^{ -0.35}_{+0.08}$ & 20 & 7.52$^{ -0.88}_{+0.46}$ & 7.71$^{ -0.74}_{+0.33}$ & 8.15$^{ -0.33}_{+0.09}$\\ 
21 & 7.71$^{ -0.91}_{+0.48}$ & 7.90$^{ -0.76}_{+0.35}$ & 8.36$^{ -0.34}_{+0.08}$ & 22 & 7.52$^{ -0.88}_{+0.47}$ & 7.70$^{ -0.74}_{+0.34}$ & 8.15$^{ -0.33}_{+0.08}$\\ 
23 & 7.62$^{ -0.89}_{+0.47}$ & 7.81$^{ -0.75}_{+0.34}$ & 8.26$^{ -0.34}_{+0.09}$ & 24 & 7.63$^{ -0.90}_{+0.48}$ & 7.81$^{ -0.75}_{+0.35}$ & 8.27$^{ -0.34}_{+0.08}$\\ 
25 & 7.65$^{ -0.90}_{+0.48}$ & 7.84$^{ -0.75}_{+0.35}$ & 8.29$^{ -0.34}_{+0.08}$ & 26 & 7.62$^{ -0.90}_{+0.47}$ & 7.81$^{ -0.75}_{+0.34}$ & 8.26$^{ -0.34}_{+0.08}$\\ 
27 & 7.60$^{ -0.90}_{+0.48}$ & 7.79$^{ -0.75}_{+0.35}$ & 8.24$^{ -0.34}_{+0.08}$ & 28 & 7.57$^{ -0.89}_{+0.47}$ & 7.75$^{ -0.74}_{+0.34}$ & 8.20$^{ -0.34}_{+0.09}$\\ 
29 & 7.63$^{ -0.90}_{+0.48}$ & 7.82$^{ -0.75}_{+0.35}$ & 8.27$^{ -0.34}_{+0.08}$ & 30 & 7.64$^{ -0.90}_{+0.47}$ & 7.83$^{ -0.75}_{+0.34}$ & 8.28$^{ -0.34}_{+0.08}$\\ 
\hline
\end{tabular}
\renewcommand{\arraystretch}{1.}
\end{center}
\caption{ \small
The total cross section for $\mu = \mt$ at Tevatron
 for $\mt$ = 171 GeV and the full set of predictions
from the MRST-2006 NNLO PDF set~\cite{Martin:2007bv}. All rates are in pb. We denote by \sigmaNLO\ \cite{Nason:1988xz,Beenakker:1989bq} the 
NLO QCD prediction by \sigmaRES\ 
the result of NLL threshold resummation \cite{Bonciani:1998vc}
and by \sigmaNNLO\ the NNLO QCD prediction based on soft
gluon approximation and exact two-loop scale dependence 
\cite{Kidonakis:2001nj}. The upper and lower indices denote
the shifts towards $\mu = 2\mt$ and $\mu=\mt/2$.}
 \end{table}

\newpage
{\footnotesize

}

\end{document}